\documentclass[journal]{IEEEtran}

\usepackage{amsthm}
\usepackage{amssymb}
\usepackage{amsfonts}
\usepackage{amsfonts,subfigure,multicol,color,verbatim,graphicx,cite,epsfig,amssymb,amsmath,cases,bm,xcolor,multirow,array}
\usepackage{algorithm,algorithmicx,algpseudocode}
\usepackage{multirow}
\usepackage{multicol}
\usepackage{supertabular}
\usepackage{array}
\usepackage{colortbl}
\usepackage{longtable}
\usepackage{subfigure}
\usepackage{dsfont}
\usepackage{url}

\usepackage{epstopdf,booktabs}

\allowdisplaybreaks[4]

\usepackage{soul} 
\usepackage{color, xcolor} 
\soulregister{\cite}7 
\soulregister{\citep}7
\soulregister{\citet}7 
\soulregister{\ref}7 
\soulregister{\pageref}7 
\soulregister{\eqref}7

\UseRawInputEncoding
\begin{document}

\title{Beam Management in Low Earth Orbit Satellite Networks with Random Traffic Arrival and Time-varying Topology}

\author{Jianfeng~Zhu,~Yaohua~Sun,~and~Mugen~Peng,~\IEEEmembership{Fellow,~IEEE}
	\thanks{Jianfeng Zhu (jianfeng@bupt.edu.cn), Yaohua Sun (sunyaohua@bupt.edu.cn), and Mugen Peng (pmg@bupt.edu.cn) are with the State Key Laboratory of Networking and Switching Technology, Beijing University of Posts and Telecommunications, Beijing 100876, China. (\bf{Corresponding author: Yaohua Sun})}}
\maketitle

\begin{abstract}
Low earth orbit (LEO) satellite communication networks have been considered as promising solutions to providing high data rate and seamless coverage, where satellite beam management plays a key role.
However, due to the limitation of beam resource, dynamic network topology, beam spectrum reuse, time-varying traffic arrival and service continuity requirement, it is challenging to effectively allocate time-frequency resource of satellite beams to multiple cells.
In this paper, aiming at reducing time-averaged  beam revisit time and mitigate inter-satellite handover, a beam management problem is formulated for dynamic LEO satellite communication networks, under inter-cell interference and network stability constraints.
Particularly, inter-cell interference constraints are further simplified into off-axis angle based constraints, which provide tractable rules for spectrum sharing between two beam cells.
To deal with the long-term performance optimization, the primal problem is transformed into a series of single epoch problems by adopting Lyapunov optimization framework.
Since the transformed problem is NP-hard, it is further divided into three subproblems, including serving beam allocation, beam service time allocation and serving satellite allocation.
With the help of conflict graphs built with off-axis angle based constraints, serving beam allocation and beam service time allocation algorithms are developed to reduce beam revisit time and cell packet queue length.
Then, we further develop a satellite-cell service relationship optimization algorithm to better adapt to dynamic network topology.
Compared with baselines, numerical results show that our proposal can reduce average beam revisit time by 20.8$\%$ and keep strong network stability
with similar inter-satellite handover frequency.
\end{abstract}


\section{Introduction}

In recent years, LEO satellite communication systems are rapidly developing, which can provide a low communication delay, high data rate and global coverage~\cite{background1}-\cite{background6}.
However, due to the non-uniform distribution of user equipments (UEs) and time-varying traffic of cells, traditional systems with fixed beam allocation suffer low resource utilization compared with systems enabling flexible time and spatial transmission\cite{hop_advantange}, namely beam hopping.
To further unleash the potential of beam hopping, effective beam management approaches play a crucial role, which faces the challenges incurred by inter-cell interference\cite{two_color}, uneven traffic distribution\cite{full_frequency_reuse2}, radio air interface delay\cite{delay}, and frequent inter-satellite handover~\cite{handover}.

Firstly, to guarantee high service quality, the interference level suffered by cells should be properly controlled.
In a full frequency reuse scenario, Lei $et$ $al.$ in~\cite{full_frequency_reuse1} proposed beam management methods to reduce co-frequency interference for a geostationary earth orbit (GEO) satellite by avoiding simultaneous transmission in adjacent cells.
Authors of~\cite{full_frequency_reuse3} proposed power allocation and many-to-many matching algorithms to make beam planning for a GEO satellite.
In addition, authors of~\cite{two_color} compared the performance of two-color and four-color schemes, where adjacent cells are allocated with beams operating with different carrier frequencies or polarization.
Although these methods can achieve significant performance improvement by interference mitigation, they ignore inter-satellite interference when multiple LEO satellites are deployed.
Actually, in some LEO constellations like Starlink, there is a high probability that multiple visible satellites are in the view of a cell simultaneously~\cite{insight_satellite,insight_satellite2}, and this can bring severe inter-satellite interference. Considering this fact, more effective beam management approaches are desired.

Secondly, to deal with spatially non-uniform UE traffic, authors of \cite{greedy1} and \cite{greedy2} leveraged greedy algorithms to make beam transmission plan.
The result showed that the designed beam hopping systems outperform random beam hopping systems in terms of throughput, delay and request satisfied ratio.
In \cite{metaheuristic_method}, the power allocation was optimized to maximize throughput and match the traffic demand based on meta-heuristic methods (e.g. genetic algorithm, simulated annealing, differential evolution and particle swarm optimization).
Moreover, literatures used genetic algorithm to match non-uniform UE traffic\cite{ga1,ga2}.
In \cite{ga1}, the transmission waiting time of data packets and network throughput were comprehensively considered in the optimization objective and power and bandwidth allocation for each beam was determined by genetic algorithm in\cite{ga2}.
Driven by the developments in artificial intelligence, authors in   \cite{AI}  propose deep reinforcement learning based beam management approach to adapt uneven traffic requests.
Although the prior works have achieved good performance in meeting traffic requests and lowering radio air interface delay, they still focus on beam management just for a static single satellite scenario without considering dynamic LEO satellites scenarios.
In addition, a cell in beam hopping systems is served by several beams operating on different frequencies and time slots in each scheduling epoch, which causes severe synchronization signaling overhead.

Thirdly, in dynamic LEO satellite networks, a satellite cannot continuously serve an earth-fixed cell due to the fast movement of satellites.
Hence, beam management approaches have to deal with inter-satellite handover.
Authors of~\cite{multi-sat1} made a beam planning for each scheduling epoch to achieve load balancing among LEO satellites, which may incur frequent inter-satellite handover.
According to~\cite{traditional_handover_1, traditional_handover_3}, traditional satellite selection criteria for inter-satellite handover
included maximal service time, highest elevation angle, and least satellite traffic load.
Since considering only a single factor was not comprehensive due to the complexity of dynamic satellite networks, some researchers further used multiple attributes for handover decision.
In \cite{multi-attribute_1}, a multi-attribute decision handover scheme was proposed based on the TOPSIS model.
The simulation results showed that multi-attribute decision based handover schemes can decrease inter-satellite handover frequency.

Although current research have achieved good performance in interference mitigation, beam hopping plan design, and inter-satellite handover, several aspects are still required to be addressed for practical LEO satellite communication networks.
Specifically, existing beam management schedules beams in per-slot manner utilizing global instantaneous channel state information, which can be infeasible in reality.
Actually, beam management decisions are usually made at ground stations and the latency incurred by collecting channel information, optimization procedure as well as
uploading scheduling results to satellites is non-negligible. Meanwhile, frequent beam management algorithm execution can put heavy computing burden on ground stations.
Hence, it is essential to manage satellite beams with a longer period and non-instantaneous but predictable information.
In addition, many existing works aim at matching the capacity of beam cells with various user traffic requests, which, however, ignores the impact of beam management on radio air interface delay and inter-satellite handover frequency. The former is related to the continuity of user service while the latter greatly affects system signalling overhead.
At last, the dynamics of user traffic arrival and network topology due to LEO satellite movement are not fully considered in the above literatures,
which makes it challenging for beam management to keep system stability and avoid inter-cell interference.

Facing the summarized deficiency, this paper formulates a novel beam management problem in LEO satellite communication networks with dynamic traffic arrival and topology, intending to lower long-term  beam revisit time and inter-satellite handover frequency.
To make the long-term optimization more tractable, the primal problem is transformed into a series of per-epoch problem by Lyapunov optimization and each epoch contains
multiple slots.
Given the high complexity of the per-epoch problem, it is further decoupled into three subproblems, including serving beam allocation, beam service time allocation and serving satellite allocation among cells.
At this time, beam management decision is made every epoch at a ground station and it mainly relies on the information of cell data queue length as well as the position information of beam cells and satellites, making our proposal more realistic for practical implementation.
The main contributions of this paper are summarized as follows.

\begin{itemize}
    \item
    A practical system model of LEO satellite communication networks is presented, which considers multi-beam LEO satellites with time varying positions, random downlink user traffic arrival, earth-fixed beam cells and beam scheduling per epoch including multiple slots.
    Based on this, a novel beam management problem is formulated, aiming at lowering long-term beam revisit time and inter-satellite handover frequency while satisfying maximal inter-cell interference constraint and data queue stability constraint of each cell.
    To make the problem more tractable, Lyapunov framework is leveraged, which decouples the long-term optimization across multiple epochs into single epoch problems.
    Moreover, the complex inter-cell interference constraints are simplified into interference constraints based on off-axis angles defined for satellite-cell pairs,
    providing an easy-to-follow rule for the ground station to judge whether two cells can share the same spectrum in advance.

	\item
    Since each single epoch problem is NP-hard with large solution space, it is further decomposed into serving beam allocation, beam service time allocation and serving satellite allocation subproblems.
    For the first subproblem, a low-complexity serving beam allocation algorithm is developed based on a conflict graph constructed with off-axis angle constraints.
    For the second subproblem, it intends to minimize the weighted sum of average beam revisit time and cell data queue length under fixed satellite-cell serving relationship by adjusting beam service starting time and service duration for each cell.
    To adapt to the dynamics of network topology, satellite-cell serving relationship is further periodically optimized by solving the third subproblem with a meta-heuristic algorithm.

    \item
    Extensive simulation is conducted to verify the effectiveness of our proposal.
    First, with the same inter-satellite handover strategy, the proposed serving beam allocation and beam service time allocation algorithms can reduce the average beam revisit time by 20.8$\%$ compared with benchmarks while resulting in shorter data queue length of cells.
    Second, by adopting the proposed serving satellite allocation algorithm, beam cells can obtain a lower beam revisit time with an affordable inter-satellite handover frequency compared with baselines.
\end{itemize}

The remainder of this paper is organized as follows.
Section~\ref{sec:system_model} describes system model and highlights several design considerations of beam management in dynamic LEO networks.
Section~\ref{sec:problem formulation} formulates the concerned dynamic beam management problem with various practical constraints. 
In Section~\ref{sec:Algorithm Framework}, corresponding algorithms are elaborated to obtain the beam management plan.
Finally, simulation results are analyzed in Section~\ref{sec:sim} followed by the conclusion in Section~\ref{sec:con}.
The notations used in this paper are summarized in Table~\ref{tab:notation}.

\begin{table}[!t]
\centering \caption{Notation}
\label{tab:notation}
\scriptsize
\begin{tabular}{l l} \toprule 
\textbf{Notation} & \textbf{Definition} \\
\midrule 
$\mathcal{S}$ & The set of satellites \\
$\mathcal{C}$ & The set of cells \\
$f$ & The index of scheduling epochs\\
$\mathcal{T}$ & The set of time slots in a scheduling epoch \\
$\mathcal{B}$ & The  set of all spot beams\\
$t_{c,f}^{\rm start}$ & The starting serving time slot of cell $c$ in  scheduling epoch $f$ \\
$t_{c,f}^{\rm end}$ & The ending serving time slot of cell $c$ in  scheduling epoch $f$ \\
$t_{c,f}$ & The beam service duration of cell $c$ in  scheduling epoch $f$ \\
$a_c$ & The mean packet arrival volume of  cell $c$ \\
$A_{c,f}$ & The number of newly arrived packets of cell $c$ in scheduling \\& epoch $f$\\
$R_{c,f}$ & The downlink rate of cell $c$ in scheduling epoch $f$. \\
$\mathcal{Q}_c$ & The virtual data queue storing the data packets of cell $c$ \\
$Q_{c,f}$ & The queue length of cell $c$ at the beginning of scheduling \\& epoch $f$ \\
$\boldsymbol{Q}_f$ & The vector of all data queue length in scheduling epoch $f$ \\
$\zeta_{b,s}$ & The  affiliation between beam $b$ and satellite $s$ \\
$\beta_{c,s,f}$ & The serving relationship between cell $c$ and satellite  $s$ in \\& scheduling epoch $f$ \\
$\alpha_{c,b,f}$ &  The serving relationship between cell $c$ and beam $b$ in  scheduling \\&  epoch $f$ \\
$\varpi_{c,c^\prime,f}$ & Indicate whether cells $c$ and $c^\prime$ have overlapping service time in \\&  scheduling epoch $f$ \\
$D_{c,f}$ & The beam revisit time of the cell $c$ in scheduling epoch $f$ \\
$D_{\rm max}$ &   The maximum  beam revisit time of cells \\
$\rho_{c,s,f}$ & The variable representing whether satellite $s$ is visible to cell $c$ \\& in scheduling epoch $f$\\
$G_{\rm user}(\theta)$ & The  receiving antenna gain on the direction of off-axis angle $\theta$\\
$G_{\rm beam}(\theta)$ &  The  transmitting antenna gain on the direction of off-axis angle $\theta$\\
$I_{c,b^\prime}$ & The strength of the interference cell $c$ suffered from beam $b^\prime$\\
$INR_{c,b}$ & The INR of cell $c$ served by beam $b$ \\
$\mathcal{B}_b$ & The  set of beams having the same frequency band as beam $b$\\
$N_{\rm noise}$ & The strength of noise signal \\
$P_{b}$ & The transmission power of beam $b$ \\
$G_{\rm max} $ & The peak gain  of satellite antenna \\
$G_{\rm min} $ & The  maximum side lobe gain of satellite antenna \\
$h_{c,b}$ &The channel gain between the center of cell $c$ and  the satellite \\& generating beam $b$ \\
$INR^{\rm th}$ & The INR threshold \\
$\overline{D}_c$ & The long-term  beam revisit time of cell $c$ \\
$\widetilde{D}_{c,f }$ & The average beam revisit time of cell $c$ from scheduling epoch \\&  $1$ to $f$ \\
$\delta_f$ & The number of cells changing the serving satellites in  the $f$-th \\& scheduling epoch \\
$\overline{\delta}$ & The long-term inter-satellite handover frequency\\
$\widetilde{\delta}_f$ & The average inter-satellite handover frequency from scheduling  \\&   epoch $1$ to $f$ \\
$L(\boldsymbol{Q}_f)$ & Lyapunov function of $\boldsymbol{Q}_f$ \\
$\Delta(\boldsymbol{Q}_f)$ & The drift of Lyapunov function $L(\boldsymbol{Q}_f)$  \\
$V$ & The parameter that controls the tradeoff between the optimization\\& objective of  problem $\boldsymbol{P}_4$ and queue length  \\
$\gamma_f$ & The objective function of problems $\boldsymbol{P}_1$ and $\boldsymbol{P}_2$ \\
$SNR_c $ & The target SNR value at the centers of cell c \\
$\mathcal{C}_{s,b,f}$ & The set of cells the satellite $s$ serves using a beam with the \\&same spectrum as beam $b$ in scheduling epoch $f$ \\
$\theta_{c,b^\prime}^{\rm tr}$ & The off-axis angle of satellite antenna between beam $b^\prime$ and \\&the center of the cell $c$ \\
$\theta_{c,b^\prime}^{\rm re}$ & The off-axis angle formed by the center of cell $c_i$ and satellites \\& generating serving beam and interference beam $b^\prime$ \\
$G^{\rm th}$ & The threshold of antenna gain attenuation \\
$\mathcal{J}_f$ & The set of interference tuples $(c,c^\prime,b,b^\prime)$\\
$v$ &The vertex in conflict graph \\
$w_v$ & The weight of vertex $v$\\
$\tau_v$ & The weight ratio of vertex $v$\\
$B_s^{\rm max}$ & The  maximum  beam number of a satellite  \\

\bottomrule 
\end{tabular}
\end{table}

\section{SYSTEM MODEL}\label{sec:system_model}

In this section, we introduce the dynamic LEO satellite network scenario and highlight the main concerns in beam management, including frequent inter-satellite handover and inter-cell interference.

\subsection{Scenario description}\label{sec:scenario}

\begin{figure}[t]
    \centering
    \begin{minipage}[t]{0.48\textwidth}
    \centering
    \subfigure[]
		{
          \includegraphics[width=3in]{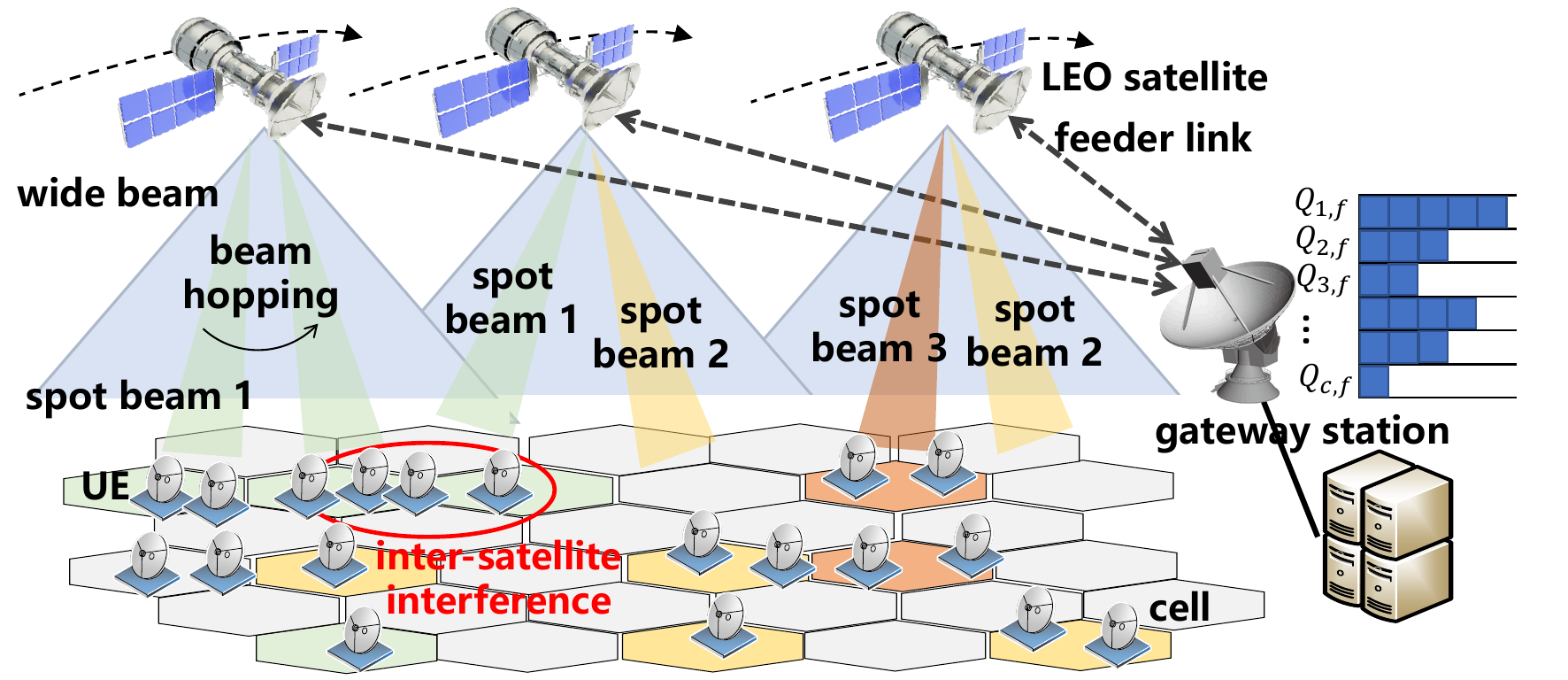}
          }
    \end{minipage}
    \begin{minipage}[t]{0.5\textwidth}
    \centering
    \subfigure[]
		{
    \includegraphics[width=3in]{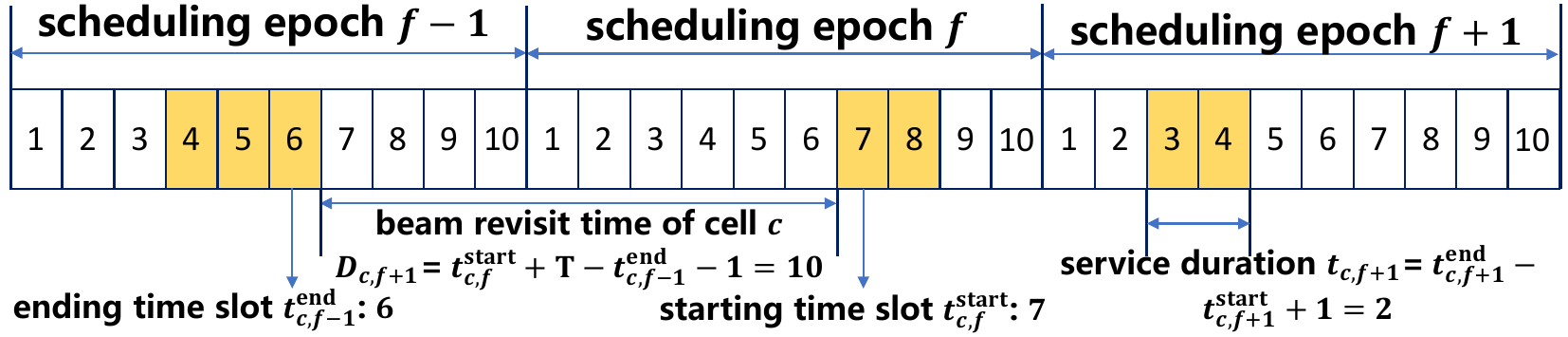}
    }
    \end{minipage}
    \caption{Figure (a) is the concerned multi-beam LEO satellite network scenario, and figure (b) provides a serving time allocation plan example of cell $c$.}\label{fig1}
\end{figure}

As shown in Fig.~\ref{fig1}(a), denote the set of LEO satellites and the set of earth-fixed cells as $\mathcal{S}=\{1,2,...,S\}$ and $\mathcal{C}=\{1,2,...,C\}$, respectively.
There is a gateway station on the ground connecting with satellites through feeder links, and it is responsible for making beam management plans for each beam scheduling epoch with $T$ time slots.
In our scenario, we only consider a local area on the surface of the Earth, where the gateway station can connect all visible satellites of cells.
Define $f=1,2,3,...$ as the index of scheduling epochs and the set of time slots in a scheduling epoch is denoted as $\mathcal T=\{1,2,...,T\}$.
Since the duration of a scheduling epoch is relatively short, each satellite's position is seen as unchanged within each scheduling epoch.

Referring to the DVB standard \cite{DVB}, LEO satellites generate several wide beams and spot beams to provide communication services.
A wide beam covers multiple cells, and the serving area of a spot beam is limited to a cell in each slot.
Moreover, wide beams deliver and receive control plane data to UEs, including initial access and inter-satellite handover data.
In addition, owing to the powerful capabilities of phased array antennas, spot beams can achieve high data rates, and beam directions can be flexibly controlled under the guidance of the beam management plan sent by the gateway station.

The operation bands of wide control beams and spot beams are assumed orthogonal.
Each spot beam of the same satellite is assumed to be allocated with a distinct band while spectrum resource is fully shared by all satellites~\cite{full_frequency_reuse2,two_color,metaheuristic_method}.
Define $\mathcal{B}=\{1,2,...,B\}$ as the set of all spot beams, and introduce a binary variable $\zeta_{b,s}$ to characterize the affiliation between beam $b$ and satellite $s$.
Specifically, $\zeta_{b,s}=1$ if the beam $b$ is generated by the satellite $s$, and $\zeta_{b,s}=0$ otherwise.
Due to large frequency offsets caused by significant relative motions between LEO satellites and cells, a beam may interfere with cells that transmit data on adjacent frequency bands.
Moreover, Doppler effect can cause inaccurate synchronization states between satellites and cells, increasing bit error rates.
To overcome these issues, we assume that satellites adopt frequency pre-compensation methods based on the velocity of satellites and the location information of satellites and cells, and then
Doppler effect can be ignored in the following analysis~\cite{38.811}.

Suppose cells obtain inconsecutive service time within a scheduling epoch.
In that case, networks must require more control signaling overhead to indicate the starting service time slot and ending slot.
Moreover, if a beam frequently adjusts its direction, it is challenging for hardware capability.
Hence, in each scheduling epoch, we assume that each cell is served by at most one spot beam in consecutive time slots to reduce the overhead.
Moreover, Fig.~\ref{fig1}(b) provides a serving time allocation plan example of cell $c$, where a scheduling epoch has ten slots, and highlighted example slots are the available transmission slots of cell $c$.
Denote the indices of the starting serving time slot and ending slot of cell $c$ in scheduling epoch $f$ as $t_{c,f}^{\rm start}$ and $t_{c,f}^{\rm end}$, respectively.
The beam service duration $t_{c,f}$ of cell $c$ in scheduling epoch $f$ is calculated as
\begin{equation}
	t_{c,f}=t_{c,f}^{\rm end}-t_{c,f}^{\rm start}+1.
	\label{eq:service duration}
\end{equation}
For example,  cell $c$ obtains two available slots in scheduling epoch $f+1$ in Fig.~\ref{fig1}(b), and its service duration $t_{c,f+1}$ is 2.

Define $A_{c,f}$ as the number of newly arrived packets of cell $c$ in scheduling epoch $f$ and assume $A_{c,f}$ follows poisson distribution with mean $a_c$.
To simplify the analysis, assume that data packets have equal size and this assumption is reasonable because the sliced packets in physical layer often have consistent length.
Note that the suffered free space path losses among the UEs in a cell served by a spot beam have only slight differences.
Hence, assuming the received power spectrum density are equal for each user in a cell and define $R_{c,f}$ as the downlink rate of cell $c$ in scheduling epoch $f$.
At the gateway station, a virtual data queue $\mathcal{Q}_c$ is maintained to store the data packets to be transmitted to cell $c$.
Define $Q_{c,f}$ as the queue length at the beginning of scheduling epoch $f$, which is updated by
\begin{equation}
	Q_{c,f+1}=\max(Q_{c,f}-t_{c,f}R_{c,f},0)+A_{c,f}.
	\label{eq:queue update}
\end{equation}

In scheduling epoch $f$, the gateway station makes a beam management plan for scheduling epoch $f+1$.
Define $\alpha_{c,b,f} \in \{0,1\}$ to indicate whether the beam $b$ allocation to cell $c$. $\alpha_{c,b,f}=1$ if cell $c$ is allocated with beam $b$ in the $f$-th scheduling epoch, and $\alpha_{c,b,f}=0$ otherwise.
Define $\beta_{c,s,f} \in \{0,1\}$ as the serving relationship between cell $c$ and satellite $s$.
$\beta_{c,s,f}=1$ means that satellite $s$ serves cell $c$ in scheduling epoch $f$, and $\beta_{c,s,f}=0$ otherwise.
Moreover, we have $\beta_{c,s,f}=\sum_{b \in \mathcal{B}}\zeta_{b,s}\alpha_{c,b,f}$.
Finally, the beam management plan for cell $c$ can be represented as a tuple $(\alpha_{c,b,f}, \beta_{c,s,f},  t_{c,f}^{\rm start}, t_{c,f}^{\rm end})$.
Based on the received plan sent by satellites via wide control beams, UEs in cell $c$ adjust operation frequency and receive data within the specified beam service time duration.
Due to the long propagation delay, the beam management plan of scheduling epoch $f$ in the actual system is determined before at least one scheduling epoch.

When making beam plan, the gateway station has to take various factors into account, including beam revisit time, network stability, inter-satellite handover frequency, and inter-cell interference.
In our scenario, beam revisit time is the interval between two consecutive transmission durations, and long  beam revisit time does harm to network stability and service continuity\cite{DVB}.
Moreover, inter-satellite handover and synchronization procedures are accomplished by wide control beams, whose delays are outside the scope of this research.
The beam revisit time $D_{c,f}$  of cell $c$ in scheduling epoch $f$ can be calculated by
\begin{equation}
	D_{c,f}= t_{c,f}^{\rm start} +T -t_{c,f-1}^{\rm end}-1,
	\label{eq: delay expectation}
\end{equation}
where $D_{c,1}=0$ and $T$ is the number of slots in each scheduling epoch.
For example, as depicted in Fig.~\ref{fig1}(b), the beam revisit time $D_{c,f}$ of cell $c$ in scheduling epoch $f$ is 10.
To achieve long-term network stability, the length of the packet queue of each cell at the gateway station should meet the following condition:
\begin{equation}
	\lim_{f \to \infty} \frac{\mathbb{E}(Q_{c,f})}{f}=0,  \;\; \forall c \in \mathcal C.
	\label{eq:queue stability}
\end{equation}
The handover among satellites and cell interference mitigation will be further discussed in the following subsections.  

\subsection{Inter-satellite handover strategy}\label{sec:switch}
Referring to \cite{time_slot_duration}, the duration of a scheduling epoch usually ranges from tens to hundreds of milliseconds, in which the locations of satellites can be assumed to be fixed.
Hence, the serving relationship between satellites and cells are seen as unchanged within an arbitrary scheduling epoch.

Define $\rho_{c,s,f}$ representing whether satellite $s$ is visible to cell $c$. $\rho_{c,s,f} =1$ if the elevation angle between the center point of cell $c$ and satellite $s$ is larger than the minimum elevation angle, and $\rho_{c,s,f} =0$ otherwise.
With satellites continuously moving, the minimum elevation angle between satellites and cells cannot be achieved, which requires inter-satellite handover.
Moreover, when $\rho_{c,s,f}\beta_{c,s,f}=1$ and $\rho_{c,s,f+1}\beta_{c,s,f+1} =0$, it means cell $c$ experiences an inter-satellite handover between scheduling epoch $f$ and $f+1$.
Given that frequent inter-satellite handover causes complex signaling procedure between UEs and satellites,
it is desired to achieve low handover frequency when designing beam management approaches.

\subsection{The modeling of inter-cell interference}\label{sec:Interference_model}

Define $\theta$ as the angle between the boresight of an interfering beam and  the direction of transmitting or receiving antenna, namely off-axis angle.
The expression of the receiving antenna gain $G_{\rm user}(\theta)$ is given by \cite{ue_Antenna}
\begin{equation}\label{rx_gain}
	G_{\rm user}(\theta)=\left\{\begin{array}{ll}
		36-25\log{\theta}, &\theta^{\rm th}\leq \theta < 44^{\circ}, \\
		\;-5 \; {\rm dBi}, &44^{\circ}\leq \theta < 75^{\circ},\\
		\;0\; {\rm dBi},&75^{\circ} \leq \theta \leq 180^{\circ},
	\end{array}\right.
\end{equation}
where $\theta^{\rm th}=\max\{1^{\circ}, 100\lambda/\kappa\}$ if $\lambda/\kappa \geq 50$, and $\theta^{\rm th}=\max\{2^{\circ}, 144(\lambda/\kappa)^{-1.09}\}$ if $\lambda/\kappa < 50$.
$\lambda$ is the wavelength of the transmitted signal and $\kappa$ is the diameter of UE antenna.
Assuming that satellite antenna arrays are distributed in the x-y plane and the weight vector of antennas are adjusted based on Chebyshev distribution, where the ratio of main lobe gain to the maximum side lobe gain is constant\cite{Optimum_array}.
To simplify the analysis, the expression of the transmitting antenna gain $G_{\rm beam}(\theta)$  can be expressed as
\begin{equation}\label{tx_gain}
	G_{\rm beam}(\theta)=\left\{\begin{array}{ll}
		G_{\rm max}, &\theta < \theta_b^{\rm 3dB}, \\
		G_{\rm min},&{\rm otherwise},
	\end{array}\right.
\end{equation}
where $\theta_b^{\rm 3dB}$ is the 3 dB angle of beam $b$, $G_{\rm max}$ and $G_{\rm min}$ are the peak gain and the maximum side lobe gain, respectively.
For a given satellite antenna size and gain ratio $G_{\rm max}/G_{\rm min}$, the weight vector of antennas can be calculated by referring to \cite{scannable_planar_arrays}.

When two satellites use beams operating on the same frequency band to serve adjacent cells, considerable inter-cell interference may occur.
The strength of downlink interfering signals is related to satellite transmission power, satellite antenna gain, the receiving gain of UE antenna, and pathloss between satellites and UEs.
Interference-to-noise ratio (INR) can be used to evaluate suffered interference level, and the INR of cell $c$ served by beam $b$ is given by
\begin{equation}
	\begin{aligned}
		INR_{c,b}=10 \log \sum_{b^\prime \in \mathcal{B}_{b} }I_{c,b^\prime}-N_{\rm noise},
	\end{aligned}
	\label{eq:INR}
\end{equation}
where $\mathcal{B}_{b}$ is the set of beams having the same frequency band as beam $b$, $N_{\rm noise}$ represents the strength of noise signal.
$I_{c,b^\prime}$ is the strength of interference from beam $b^\prime$, which is given by
\begin{equation}
 I_{c,b^\prime}= P_{b^\prime} G_{\rm beam}(\theta_{c,b^\prime}^{\rm tr}) G_{\rm user}(\theta_{c,b^\prime}^{\rm re}) h_{c,b^\prime}
\end{equation}
where $P_{b^\prime}$ is the transmission power of beam $b^\prime$.
$G_{\rm beam}(\theta_{c,b^\prime}^{\rm tr})$ is the transmitting gain of beam $b$ on off-axis angle $\theta_{c,b^\prime}^{\rm tr}$,
$G_{\rm user}(\theta_{c,b^\prime}^{\rm re})$ is the receiving gain of UEs on the off-axis angle $\theta_{c,b^\prime}^{\rm re}$ in cell $c$.
Off-axis angles $\theta_{c,b^\prime}^{\rm tr}$ and $\theta_{c,b^\prime}^{\rm re}$ are determined by the location of cell $c$ and the direction of beams $b^\prime$ and $b$\cite{off-axis}.
$h_{c,b^\prime}$ is the channel gain between the center of cell $c$ and the satellite generating beam $b^\prime$, which depends on free space loss, rain attenuation, atmospheric gas attenuation, and cloud and fog attenuation\cite{ITU}.

\section{Problem Formulation And Transformation}\label{sec:problem formulation}

In this section, we first formulate several realistic constraints and the concerned beam management problem for dynamic LEO satellite networks.
Then, to make the problem more tractable, it is further transformed with the help of Lyapunov drift rule.
Considering that INR expression involves future channel state information, which cannot be obtained by the gateway station when making the beam management plan for the incoming epoch, INR constraints are hence simplified into intuitive space angle constraints based on the propagation loss model proposed by ITU\cite{ITU}.

\subsection{Problem formulation }\label{sec:Problem Formulation}

\subsubsection{Serving satellite allocation constraints}

When making a beam management plan, the gateway station allocates each cell with at most one satellite that satisfies the minimum elevation angle requirement.
Then, we have
\begin{equation}
	\sum_{s=1}^S \rho_{c,s,f} \beta_{c,s,f} =1, \;\; \forall c \in \mathcal C,  \forall f,
	\label{eq:satellite constraints}
\end{equation}
where $ \rho_{c,s,f}=1$ represents the elevation angle between the center of cell $c$ and satellite $s$ exceeding the minimum threshold.

\subsubsection{Beam allocation constraints}
To avoid frequent inter-beam handover, a cell is always served by the same beam in a scheduling epoch, which is expressed as
\begin{equation}\label{slot beam index}
 \sum_{b \in \mathcal{B} } \alpha_{c,b,f} \leq 1,\;\; \forall c,f.
\end{equation}

\subsubsection{Maximal INR constraints}

Define $INR^{\rm th}$ as the INR threshold, and then we have the following INR constraints based on (\ref{eq:INR}): 
\begin{equation}
	\sum_{b \in \mathcal{B}_{b} } \alpha_{c,b,f}INR_{c,b} \leq INR^{\rm th},  \; \; \forall c \in \mathcal C,  \forall b \in \mathcal B, \forall f.
	\label{eq:INR constraints}
\end{equation}

\subsubsection{The objective function of beam management}

Denote the average beam revisit time of cell $c$ as $\overline{D}_c$, which is calculated by
\begin{equation}
	\overline{D}_c = \lim_{f^\prime \to \infty}  \frac{1}{f^\prime} \sum_{f=1}^{f^\prime} D_{c,f}.
	\label{eq: delay expectation}
\end{equation}

Denote the number of cells changing the serving satellites between the $f-1$-th and the $f$-th scheduling epochs as $\delta_f$, which is given by
\begin{equation}
	\delta_f = \sum_{c=1}^C  (1-\sum_{s=1}^{S}\beta_{c,s,f-1}\beta_{c,s,f}),
\end{equation}
and $\delta_1=0$.
Then, the average number of cells experiencing inter-satellite handover is given by
\begin{equation}
	\overline{\delta} = \lim_{f^\prime \to \infty}  {\frac{1}{f^\prime}\sum_{f=1}^{f^\prime}} \delta_f.
	\label{eq: handover satellite expectation}
\end{equation}

Finally, we define the following long-term objective for our beam management to lower beam revisit time and inter-satellite handover frequency, which is
\begin{equation}
	\sum_{c=1}^C   \frac{\overline{D}_c }{1+C-\overline{\delta}},
	\label{eq: po_objective}
\end{equation}
where the denominator of (\ref{eq: po_objective}) is constructed by the difference between the total number of cells and the average number of cells that perform handover in each scheduling epoch.
Hence, larger denominator indicates a lower inter-satellite handover frequency.
The numerator of (\ref{eq: po_objective}) represents the average sum of  beam revisit time of cells.

\subsubsection{Problem formulation}

With the above objective and design constraints, the concerned beam management problem in dynamic LEO satellite networks is formulated as follows
\begin{align}
	&\boldsymbol{P_0}: \min\limits_{\{\alpha_{c,b,f},\beta_{c,s,f},t_{c,f}^{\rm start},t_{c,f}^{\rm end} |\forall c,s,b,f\}}  \quad \sum_{c=1}^C   \frac{\overline{D}_c }{1+C-\overline{\delta}}   \label{p0} \\
	s.t. \;& D_{c,f}<D_{\rm max},  \;\; \forall c \in \mathcal C, \forall f  \tag{\ref{p0}{a}} \label{p0a},\\
	&   0 \leq t_{c,f}^{\rm start} \leq t_{c,f}^{\rm end} \leq T, \;\; \forall c \in \mathcal C, \forall f \tag{\ref{p0}{b}} \label{p0b},\\
	& \alpha_{c,b,f},\beta_{c,s,f} \in \{0,1\}, \; \; \forall c, s,   b,  f, \tag{\ref{p0}{c}} \label{p0c}\\
	& (\ref{eq:queue stability}),(\ref{eq:satellite constraints})-(\ref{eq:INR constraints}).   \notag
\end{align}
Constraint (\ref{p0}{a}) restricts the maximum beam revisit time of cells to no more than $D_{\rm max}$.
Constraint (\ref{p0}{b}) indicates that the index of service ending slot $t_{c,f}^{\rm end}$ must be larger that or equal to the index of service starting slot $t_{c,f}^{\rm start}$.
Constraint (\ref{p0}{c}) means variables $\alpha_{c,b,f}$ and $\beta_{c,s,f}$ are binary, and other constraints have been introduced as aforementioned.

\subsection{Problem transformation }\label{sec:Problem Transformation}

It can be seen that the objective of problem $\boldsymbol{P_0}$ and network stability constraint (\ref{eq:queue stability}) involve time averaged statistical values, which imposes significant challenge in directly solving the problem.
In this part, we transform problem $\boldsymbol{P_0}$ into a tractable form with the help of Lyapunov drift rule.

First, two metrics are defined as follows:
\begin{equation}
	\widetilde{D}_{c,f }=\frac{1}{f} \sum_{f^\prime=1}^{f} D_{c,f^\prime}, \; \; \; \widetilde{\delta}_f=\frac{1}{f} \sum_{f^\prime=1}^{f} \delta_{f^\prime}.
	\label{eq:average delay cal}
\end{equation}
Denote the feasible solution set to problem $\boldsymbol{P_0}$ as $\mathcal{P}_0$ that is a set of tuple $\{\alpha_{c,b,f},\beta_{c,s,f},t_{c,f}^{\rm start},t_{c,f}^{\rm end}\}$.
The components of each solution in $\mathcal{P}_0$ corresponding to scheduling epoch $f$ is denoted by $p_f$.
Note that $\overline{D}_{c,f} \in (0,D_{\rm max})$ and $\overline{\delta} \in [0,C]$, and then we have
\begin{equation}
	\left\{\begin{array}{ll}
		0 \leq \mathbb{E}(\widetilde{D}_{c,f}  | p_f )  \leq D_{\rm max}, \\
		1 \leq \mathbb{E}(1+C- \widetilde{\delta}_f | p_f )  \leq  C+1,
	\end{array}\right.
\end{equation}
and
\begin{equation}
	\left\{\begin{array}{ll}
		\mathbb{E}((\widetilde{D}_{c,f}  | p_f )^2 | p_f)  \leq D_{\rm max}^2, \\
		\mathbb{E}((1+C- \widetilde{\delta}_f )^2| p_f )  \leq (C+1)^2,
	\end{array}\right.
\end{equation}
where $C$ is the total number of cells and $D_{\rm max}$ is the maximum beam revisit time.

Define $\boldsymbol{Q}_f = (Q_{1,f},...,Q_{C,f})$ as the vector of all data queue, and then the Lyapunov function $L(\boldsymbol{Q}_f)$ can be constructed as $L(\boldsymbol{Q}_f)=\frac{1}{2} \sum_{c=1}^C Q_{c,f}^2$.
Thus, the drift of Lyapunov function $L(\boldsymbol{Q}_f)$ can be obtained as \cite{boundedness}
\begin{equation}
	\Delta(\boldsymbol{Q}_f) \triangleq L(\boldsymbol{Q}_{f+1})-L(\boldsymbol{Q}_f).
\end{equation}

Based on expression  (\ref{eq:queue update}), under any beam scheduling decision $p_f $, we have \cite{Queueing_object1}
\begin{equation}
\begin{aligned}
	\mathbb{E}[\Delta(\boldsymbol{Q}_f)|p_f] &\leq \frac{1}{2} \sum_{c=1}^C (A_{c,f}-t_{c,f}R_{c,f})^2 \\ &+\sum_{c=1}^C \mathbb{E}[(A_{c,f}-t_{c,f}R_{c,f})Q_{c,f}|p_f],
\end{aligned}
\label{eq:xxx}
\end{equation}

Define auxiliary variable $\eta_{c,f}$  for cell $c$ as follows:
\begin{equation}
	\eta_{c,f}=\frac{ \sum_{f^\prime=1}^{f-1} D_{c,f^\prime}}{ \sum_{f^\prime=1}^{f-1} (1+C-\delta_f^\prime)},
	\label{eta}
\end{equation}
and $\eta_{c,1}=0$.
Based on the boundedness analysis \cite{boundedness} and Lebesgue dominated convergence theorem\cite{Queueing_book}, we can apply Lyapunov drift to transform the primal problem $\boldsymbol{P}_0$.
Then, by following literature~\cite{boundedness,Queueing_object,Queueing_object1}, the objective in primal problem $\boldsymbol{P}_0$ can be transformed into
\begin{equation}
	\min \mathbb{E}[\Delta(\boldsymbol{Q}_f)+V\sum_{c=1}^C (\widetilde{D}_{c,f}- \eta_{c,f}(1+C-\widetilde{\delta}_f) )|p_f],
\end{equation}
where $V > 0$ is a parameter that controls the tradeoff between the optimization objective of primal problem $\boldsymbol{P_0}$ and queue length.
According to (\ref{eq:xxx}), we have \cite{boundedness}
\begin{equation}\label{fitness}
\begin{aligned}
&	\mathbb{E}[\Delta(\boldsymbol{Q}_f)+\sum_{c=1}^C V(\widetilde{D}_{c,f}- \eta_{c,f}(1+C-\widetilde{\delta}_f) )|p_f] \\
& \leq  \overline B+\sum_{c=1}^C \mathbb{E}[(A_{c,f}-t_{c,f}R_{c,f})Q_{c,f}|p_f] \\
&+ \mathbb{E}[V\sum_{c=1}^C (\widetilde{D}_{c,f}- \eta_{c,f}(1+C-\widetilde{\delta}_f) )|p_f]
\end{aligned}
\end{equation}
where $\overline B $ is a constant and $\overline B \geq \frac{1}{2} \sum_{c=1}^C (A_{c,f}-t_{c,f}R_{c,f})^2 $.
Recalling that $A_{c,f}$ is the number of newly arrived packets of cell $c$ in the $f$-th scheduling epoch, the value of term $Q_{c,f}A_{c,f}$ can be regarded as a constant.
Moreover, according to the finiteness of inequality (\ref{fitness}), $\exists \Omega$ and it satisfies $\Delta(\boldsymbol{Q}_f) \leq \Omega$.
Further, we have $\mathbb{E}[\Delta(\boldsymbol{Q}_f)]-\mathbb{E}[\Delta(\boldsymbol{Q}_0)]\leq f\Omega$ \cite{delay}.
Then, we obtain that $\mathbb{E}[(Q_{c,f})^2] \leq 2f\Omega + \mathbb{E}[(Q_{c,0})^2] $.
Since $\mathbb{E}[(Q_{c,f})^2] \geq \mathbb{E}^2[Q_{c,f}]$,
we have
\begin{equation}\label{stability}
	\lim_{f \to \infty} \frac{ \mathbb{E}[Q_{c,f}]}{f} \leq  \lim_{f \to \infty} \frac{\sqrt{2f\Omega + \mathbb{E}[(Q_{c,0})^2]}}{f} =0.
\end{equation}

Based on the above analysis, primal problem $\boldsymbol{P}_0$ can be minimized by designing a beam management algorithm that minimizes the following objective in each scheduling epoch:
\begin{equation} \label{eq:gamma}
	\begin{aligned}
		\gamma_f=&\sum_{c=1}^C V(\widetilde{D}_{c,f}- \eta_{c,f}(1+C-\widetilde{\delta}_f) )\\& -R_{c,f}Q_{c,f}(t_{c,f}^{\rm end}-t_{c,f}^{\rm start}+1).
	\end{aligned}
\end{equation}
In addition, minimizing objective $\gamma_f$ can guarantee the stability of LEO communication systems according to  inequality (\ref{stability}) \cite{Queueing_object}.
Finally, the primal problem $\boldsymbol{P_0}$ is replaced by problem $\boldsymbol{P_1}$ in each scheduling epoch $f$ as follows:
\begin{align}
	\boldsymbol{P_1}: & \min\limits_{\{\alpha_{c,b,f},\beta_{c,s,f},t_{c,f}^{\rm start},t_{c,f}^{\rm end} |\forall c,s,b  \}} \quad	\gamma_f   \label{p1}\\
	s.t. \;&(\ref{eq:satellite constraints})-(\ref{eq:INR constraints}), (\ref{p0}{a})-(\ref{p0}{c}),  (\ref{eq:gamma}).  \notag
\end{align}

\subsection{The simplification of INR constraints }\label{sec:PInterference Cancellation}

In problem $\boldsymbol{P_1}$, the transmission power in INR constraint (\ref{eq:INR constraints}) depends on future channel conditions, which makes beam management design intractable.
Therefore, we try to convert the INR constraint into space angle related constraints that only rely on satellite and cell positions.
With such constraints, the gateway station can easily identify whether two cells can be allocated with the same frequency band.
At the beginning, we first consider the scenario containing two cells and two satellites and then extend the analysis to multi-satellite scenario.

\begin{figure}[!t]
	\center
	\includegraphics[width=2.4in]{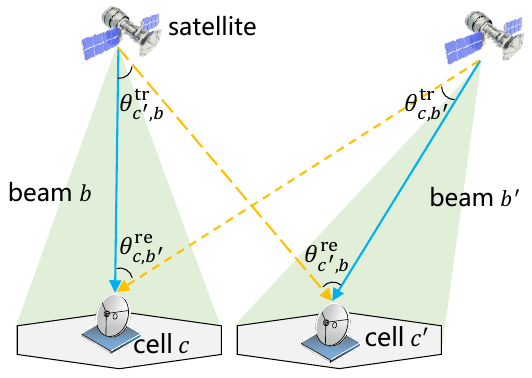}
	\caption{The interference scenario between two satellite-cell pairs.}\label{fig2}
\end{figure}

As indicated in Fig.~\ref{fig2}, cell $c$ and $c^\prime$ are served by two beams $b$ and $b^\prime$ with the same operation frequency, respectively.
Denote the transmission power of two beams as $P_{b}$ and $P_{b^\prime}$, respectively.
Hence, the target signal-to-noise ratio (SNR) values at the centers of cell $c$ and $c^\prime$ are given by
\begin{equation}
	\left\{
	\begin{array}{lr}
		SNR_{c} =10\log{P_{b}G_{\rm beam}(0)G_{\rm user}(0)h_{c,b}-N_{\rm noise}}, \\
		SNR_{c^\prime} =10\log{P_{b^\prime}G_{\rm beam}(0)G_{\rm user}(0)h_{c^\prime,b^\prime}- N_{\rm noise}},
	\end{array}
	\right.
	\label{cell snr}
\end{equation}
where $G_{\rm user}(.)$ and $G_{\rm beam}(.)$  are the expression of the receiving antenna gain and  transmitting antenna gain in (\ref{rx_gain}) and (\ref{tx_gain}), respectively.
$h_{c,b}$ is the channel gain between the satellite generating beam $b$ and the center of cell $c$, $h_{c^\prime,b^\prime}$ is the channel gain between the satellite generating beam $b^\prime$ and the center of cell $c^\prime$.
Meanwhile, the INR of cell $c$ is given by
\begin{equation} \label{cell ci inr}
	\begin{aligned}
		INR_{c,b} =10\log{I_{c,b^\prime}}-N_{\rm noise} \leq INR^{\rm th},
	\end{aligned}
\end{equation}
where $I_{c,b^\prime}=P_{b^\prime}G_{\rm beam}(\theta_{c,b^\prime}^{\rm tr})G_{\rm user}(\theta_{c,b^\prime}^{\rm re})h_{c,b^\prime}$,
$G_{\rm beam}(\theta_{c,b^\prime}^{\rm tr})$ is the transmission gain of satellite antenna on the direction of off-axis angle $\theta_{c,b^\prime}^{\rm tr}$,
$G_{\rm user}(\theta_{c,b^\prime}^{\rm re})$ represents the receiving gain of UE receiving antenna on the direction of off-axis angle $\theta_{c,b^\prime}^{\rm re}$, and $INR^{\rm th}$ is the INR threshold.

Based on the second term of (\ref{cell snr}) and  the inequation (\ref{cell ci inr}), 
we can obtain
\begin{equation}
	10\log\frac{G_{\rm beam}(\theta_{c,b^\prime}^{\rm tr})G_{\rm user}(\theta_{c,b^\prime}^{\rm re})h_{c,b^\prime}}{G_{\rm beam}(0)G_{\rm user}(0)h_{c^\prime,b^\prime}} \leq INR^{\rm th}-SNR_{c^\prime}.
	\label{eq:relationship}
\end{equation}

In order to guarantee that INR constraint (\ref{eq:INR constraints}) always holds, we define the minimum ratio $h_{\rm min}$ between $h_{c,b^\prime}$ and $h_{c^\prime,b^\prime}$ and consider the following constraint on the basis of inequality (\ref{eq:relationship}).
\begin{equation}
	10\log\frac{G_{\rm beam}(\theta_{c,b^\prime}^{\rm tr})G_{\rm user}(\theta_{c,b^\prime}^{\rm re})}{G_{\rm beam}(0)G_{\rm user}(0)h_{\rm min}} \leq INR^{\rm th}-SNR_{c^\prime}.
	\label{eq:relationship1}
\end{equation}

Then, for the scenario with multiple satellites, we obtain the following interference constraint for cell $c$:
\begin{equation}
	10\log\frac{S_{\rm max}G_{\rm beam}(\theta_{c,b^\prime}^{\rm tr})G_{\rm user}(\theta_{c,b^\prime}^{\rm re})}{G_{\rm beam}(0)G_{\rm user}(0)h_{\rm min}} \leq INR^{\rm th}-SNR_{c^\prime},
	\label{eq:relationship3}
\end{equation}
where $S_{\rm max}$ is the maximal number of satellites that are visible to cell $c$ simultaneously.

Introduce an auxiliary variable $\varpi_{c,c^\prime,f} \in \{0,1\}$ to indicate whether cells $c$ and $c^\prime$ have overlapping service time in scheduling epoch $f$.
$\varpi_{c,c^\prime,f} =1$ if the service time of cells $c$ and $c^\prime$ overlaps, and $\varpi_{c,c^\prime,f} =0$ otherwise.
Further, constraint (\ref{eq:relationship3}) can be transformed into the following form:	
\begin{equation}
\begin{aligned}
		&G_{\rm beam}(\theta_{c,b^\prime}^{\rm tr})G_{\rm user}(\theta_{c,b^\prime}^{\rm re}) < G^{\rm th}, \;  \forall c, c^\prime \in \mathcal C, \\ &  \forall f, if \; \alpha_{c,b,f}+\alpha_{c^\prime,b^\prime,f}+\varpi_{c,c^\prime,f} =3,
	\label{eq:angle_constraint1}
\end{aligned}
\end{equation}
where $G^{\rm th}= 10^{(INR^{\rm th}-SNR_{c^\prime}-10\log{(\frac{S_{\rm max}}{G_{\rm beam}(0)G_{\rm user}(0)h_{\rm min}})})/10}$.

Clearly, if condition (\ref{eq:angle_constraint1}) holds for all cells, INR constraint is met for each cell.
Meanwhile  condition (\ref{eq:angle_constraint1}) only relies the locations of cells and satellites.
Based on these location information, a set $\mathcal{J}_f$ can be generated by the gateway station based on (\ref{eq:angle_constraint1}).
Specifically, if $(c,b,c^\prime,b^\prime)$ belongs to $\mathcal{J}_f$, the INR constraint of cell $c$ or $c^\prime$ will be violated if cell $c$ is served by beam $b$ while cell $c^\prime$ is served by beam $b^\prime$.
Moreover, beams $b$ and $b^\prime$ operate on the same frequency band.
After defining $\mathcal{J}_f$, INR constraint (\ref{eq:angle_constraint1}) is equivalent to
\begin{equation}
 \alpha_{c,b,f}+\alpha_{c^\prime,b^\prime,f}+\varpi_{c,c^\prime,f} <3, \forall (c,b,c^\prime,b^\prime) \in \mathcal {J}_f.
	\label{eq:angle constraint2}
\end{equation}

Finally, the dynamic beam management problem in scheduling epoch $f$ is transformed as follows:
\begin{align}
	\boldsymbol{P_2}: & \min\limits_{\{\alpha_{c,b,f},\beta_{c,s,f},t_{c,f}^{\rm start},t_{c,f}^{\rm end} |\forall c,s,b \}}  \quad	\gamma_f   \label{p1}\\
	s.t.\;& (\ref{eq:satellite constraints}),(\ref{slot beam index}),  (\ref{p0}{a})-(\ref{p0}{c}), (\ref{eq:gamma}),(\ref{eq:angle constraint2}).  \notag
\end{align}

\section{Problem Decomposition and Beam Management Algorithm Design}\label{sec:Algorithm Framework}

\begin{figure}[hbtp]
	\center
	\includegraphics[width=3.3in]{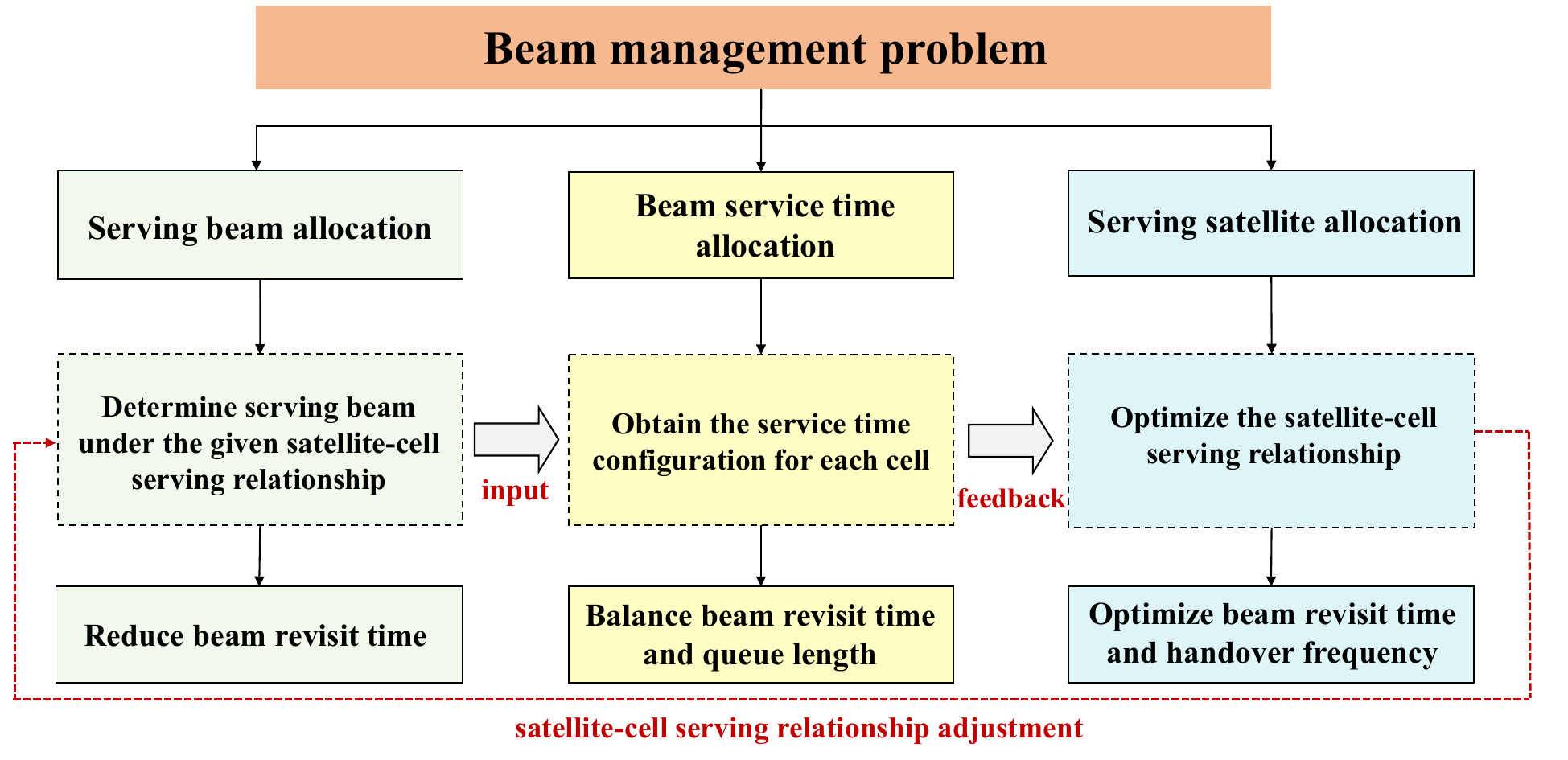}
	\caption{Beam management problem decomposition.}\label{fig15}
\end{figure}

When the number of visible LEO satellites in the concerned area is one, problem $\boldsymbol{P_2}$ reduces to a Vehicle Routing Problem, which is NP-hard~\cite{NP-hard}.
Therefore, problem $\boldsymbol{P_2}$ is also an NP-hard problem, and it is difficult to obtain the optimal solution for actual networks under affordable complexity. 
To make the problem more tractable, this section further decouples problem $\boldsymbol{P_2}$ into three closely related subproblems as shown in Fig.~\ref{fig15}, namely
serving beam allocation problem, beam service time allocation problem, and serving satellite allocation problem.
The decomposition reasons are summarized as follows.

Firstly, satellite-cell serving relationships are tightly coupled with beam-cell serving relationships due to complex inter-beam interference.
Moreover, if the proposed serving satellite allocation algorithm is executed in each scheduling epoch, it is inevitable for cells to undergo high inter-satellite handover frequency.
Hence, the serving satellite allocation problem is first decomposed.
In addition, the concerned problem has massive serving beam allocation and serving time allocation variables, which are also tightly coupled.
In this case, the computational complexity of branch and bound and Benders decomposition methods to solve problem is unacceptable for practical applications \cite{solver}.
Then, we further decompose the serving beam allocation problem and the beam service time allocation problem from problem $\boldsymbol{P_2}$.

In each scheduling epoch, if serving satellite allocation algorithm is not performed, the satellite-cell relationships $\beta_{c,s,f}$ remain the same in the last epoch.
In this case, LEO satellite networks first execute serving beam allocation algorithm and find the set of $(\alpha_{c,b,f},t_{c,f}^{\rm start})$ to reduce beam revisit time.
Subsequently, beam service time allocation algorithm is performed based on the output of the serving beam allocation algorithm, aiming at balancing beam revisit time and queue length.
Specifically, the final beam management plan for this scheduling epoch is outputted by beam service time allocation algorithm when satellite-cell relationships $\beta_{c,s,f}$ are fixed.
If serving satellite allocation algorithm is carried out, it will invoke serving beam allocation and beam service time allocation algorithms to update $\beta_{c,s,f}$ and  obtain final beam management plan.

\subsection{Serving beam allocation problem and algorithm design}\label{sec:subproblem 1}

When fixing satellite-cell serving relationship $\beta_{c,s,f}$ and beam service duration $t_{c,f}$, terms $\eta_{c,f}(1+C-\widetilde{\delta}_f)$ and $R_{c,f}Q_{c,f}(t_{c,f}^{\rm end}-t_{c,f}^{\rm start}+1)$ in (\ref{eq:gamma}) are constant in scheduling epoch $f$.
Thus, serving beam allocation problem in scheduling epoch $f$ is written as follows.
\begin{align}
	\boldsymbol{P_3}:	& \min\limits_{\{\alpha_{c,b,f},t_{c,f}^{\rm start} |\forall c,b \}}  \quad	 \sum_{c=1}^C \widetilde{D}_{c,f}	 \label{BCA}\\
	s.t.\;& (\ref{eq:service duration}), (\ref{slot beam index}), (\ref{p0}{a})-(\ref{p0}{c}),  (\ref{eq:angle constraint2}).  \notag
\end{align}

Problem $\boldsymbol{P_3}$ aims to find serving beam allocation $\alpha_{c,b,f}$ and beam service starting time slot $t_{c,f}^{\rm start}$ for each cell to minimize average beam revisit time.
To solve problem $\boldsymbol{P_3}$, a weighted conflict graph is first built, and vertex set and edge set in this graph are denoted by $\mathcal{V}$ and $\mathcal{E}$, respectively.
Specifically, each vertex $v \in \mathcal{V}$ represents a feasible time-frequency allocation decision tuple $(\alpha_{c,b,f},t_{c,f}^{\rm start})$ for a cell $c$, which
means that cell $c$ obtains service starting from time slot $t_{c,f}^{\rm start}$ by beam $b$ in scheduling epoch $f$.
Moreover, the number of vertices equals the number of time-frequency allocation decision tuples of all cells, which ensures one-to-one mapping between a vertex $v$ and $(\alpha_{c,b,f},t_{c,f}^{\rm start})$.
In addition, each vertex $v$ has its own weight and the weight $w_{v}$ corresponding to tuple $(\alpha_{c,b,f},t_{c,f}^{\rm start})$ of cell $c$ is set as
\begin{equation}
\begin{aligned}
	& w_{v} =D_{\rm max}-\widetilde{D}_{c,f} \\ &=D_{\rm max}- \frac{(f-1)\widetilde{D}_{c,f-1} + (T+t_{c,f}^{\rm start} - t_{c,f-1}^{\rm end}-1) }{f},
	\label{eq:vertex weight}
\end{aligned}
\end{equation}
where $T$ is the total number of time slots in a scheduling epoch.
Meanwhile, there exists an edge between two vertices if one of the following cases occurs.
\begin{itemize}		
	\item Two vertices represent the same cell.
	\item Two vertices represent two cells served by two beams with the same spectrum, and their beam service time overlaps.
	\item Constraint (\ref{eq:angle constraint2}) is violated.
\end{itemize}

\begin{figure}[hbtp]
	\center
	\includegraphics[width=2.8in]{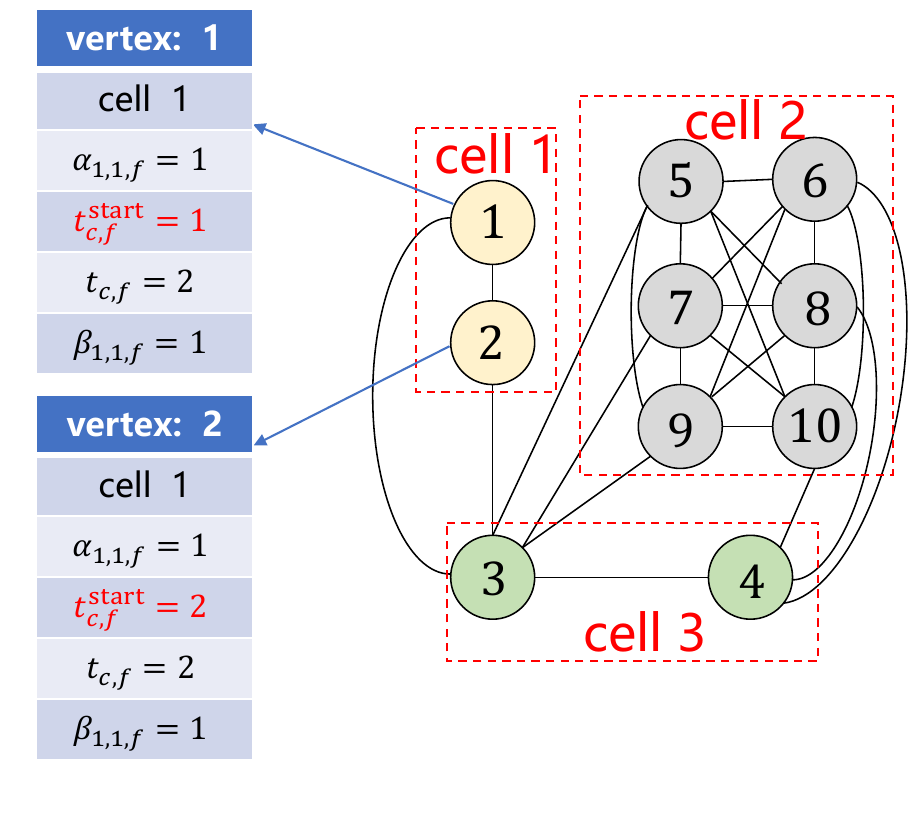}
	\caption{An example of constructed conflict graph. }\label{fig3}
\end{figure}

Fig.~\ref{fig3} illustrates an example of a conflict graph for scheduling epoch $f$,
in which cell 1 is served by satellite 1 with one beam, and satellite 2 with two beams serves cell 2 and 3.
In addition, the beam of satellite 1 operates at the same frequency band with one of beam of satellite 2.
The service duration of 3 cells is 2, 1, and 3 slots, respectively.
Hence, there are 2, 6 and 2 vertices corresponding to cell 1, 2 and 3.
Suppose that only cell 1 and cell 3 have intolerable co-frequency interference if they are served by beams with the same frequency band.
Then, two vertices of cell 1 connect to all vertices of cell 3.

With conflict graph, problem $\boldsymbol{P_3}$ can be transformed into a weighted maximum independent set problem, where a vertex subset $\mathcal{V}^\prime$ is selected such that any two vertices in $\mathcal{V}^\prime$ are not connected and the sum of weights of all vertices in $\mathcal{V}^\prime$ achieve the maximum~\cite{NP-hard1}.
To obtain the maximum weighted independent set $\mathcal{V}^\prime$ in conflict graphs, we propose a low complexity greedy search algorithm shown in Algorithm~\ref{alg:BA_Algorithm}.
At first, the weight ratio $\tau_v$ for each vertex $v$ is calculated by
$\tau_v =\frac{w_v}{w_v+ \sum_{v^\prime \in \mathcal{V}_v}w_{v^\prime}}$, where $\mathcal{V}_{v}$ is the set of interconnected vertices of vertex $v$ in conflict graph (line 2).
Next, according to the descending order of weight ratios, vertices are sorted and the index of each vertex after sorting is recorded (line 3).
Subsequently, we search each optional vertex in descending order of vertex index in a loop and add the current vertex to vertex set $\mathcal{V}^\prime$ if this vertex and all vertices in $\mathcal{V}^\prime$ are not connected.
Meanwhile, we set an inaccessible state for all vertices directly connected to this vertex (line 4-13), and
the allocated time-frequency resource for the cell corresponding to the vertex is set as being occupied.
Note that the number of chosen vertices may be less than the number of cells.
Therefore, the cell without any allocated resource is assigned with a tuple of $(\alpha_{c,b,f}, t_{c,f}^{\rm start})$ if constraint (\ref{eq:angle constraint2}) holds (line 14-19).
In this case, the service duration of that cell is one time slot.
Finally, we obtain configuration tuples $(\alpha_{c,b,f}, t_{c,f}^{\rm start},t_{c,f})$ for all cells (line 20).

\begin{algorithm} [htbp]
\begin{algorithmic}[1]
    \footnotesize
	\caption{Serving Beam Allocation Algorithm}
	\label{alg:BA_Algorithm}
	\State Input: $\beta_{c,s,f}$, $t_{c,f}$, $T$, $D_{\rm max}$, $\widetilde{D}_{c,f-1}$, $t_{c,f-1}^{\rm end}$, $R_{c,f}$, $\mathcal{V}^\prime= \emptyset$, $\mathcal Result = \emptyset$, $\boldsymbol{f_c}=\boldsymbol{0}_{1 \times C}$, $\boldsymbol{f_v}=\boldsymbol{0}_{1 \times |\mathcal{V}|}$.
	\State Construct a conflict graph, calculate the weight radio $\tau_{v}$ for all vertices.
	\State According to descending order of weight radio, sort vertices and record the new index of each vertex.
	\For{$k=1:|\mathcal{V}|$}
	\If{$ f(k)==0  $}
	\State Find the vertex $v$ corresponding to index $k$.
	\If{$\mathcal{V}^\prime == \emptyset$   or $ (v, v^\prime)  \not\in \mathcal{E},\; \forall v^\prime \in \mathcal{V}^\prime $}
	\State Record configuration tuple $(\alpha_{c,b,f},t_{c,f}^{\rm end},t_{c,f})$ based on (\ref{eq:service duration}) and (\ref{slot beam index}) corresponding to vertex $v$: $\mathcal Result \gets \mathcal Result \cup \{(\alpha_{c,b,f},t_{c,f}^{\rm end},t_{c,f})\}$.
	\State $\mathcal{V}^\prime\gets \mathcal{V}^\prime \cup \{v\}$, $\boldsymbol{f_c}(c)=1$,  where $t_{c,f}^{\rm start}  \leq t \leq t_{c,f}^{\rm start}+t_{c,f}-1$.
	\State Set an inaccessible state $\boldsymbol{f_v}(k)=1$ for all vertex directly connected to vertex $v$ and record the allocated time-frequency resource.
	\EndIf
	\EndIf
	\EndFor
	\For{$c=1:C$}
	\If{$ \boldsymbol{f_c}(c)==0 $}
	\State Find an unallocated time-frequency resource with $\boldsymbol{f_r}(b,t)=0$ for cell $c$, and the constraint (\ref{eq:angle constraint2}) hold.
	\State  $\alpha_{c,b,f} \gets 1$, $t_{c,f}^{\rm start}  \gets t$, $\mathcal Result \gets \mathcal Result \cup {\{(\alpha_{c,b,f},t_{c,f}^{\rm start},1)\}}$.
	\EndIf
	\EndFor
	\State Output:  $\mathcal Result$.
	\end{algorithmic}
\end{algorithm}

\subsection{Beam service time allocation problem and algorithm design}\label{sec:subproblem 2}

In this subsection, beam service time allocation to each cell is further optimized based on pre-fixed serving satellite allocation $\beta_{c,s,f}$, serving beam allocation $\alpha_{c,b,f}$ and initial service starting time slot index $t_{c,f}^{\rm start}$, aiming to balance beam revisit time and queue length.
According to (\ref{eq: delay expectation}) and (\ref{eq:average delay cal}), $\widetilde{D}_{c,f}$ can be rewritten as
\begin{equation}
	\begin{aligned}
		\widetilde{D}_{c,f} =& \frac{1}{f} \sum_{f^\prime=1}^f D_{c,f^\prime}\\ =&\frac{T+t_{c,f}^{\rm start}-1- t_{c,f-1}^{\rm end}+(f-1)\widetilde{D}_{c,f-1}}{f}.
	\end{aligned}
	\label{eq:average delay trans}
\end{equation}

Since $T- t_{c,f-1}^{\rm end}-1+(f-1)\widetilde{D}_{c,f-1}$ in (\ref{eq:average delay trans}) and $\eta_{c,f}(1+C-\widetilde{\delta}_f)$ in (\ref{eq:gamma}) are irrelevant to$\alpha_{c,b,f}$, $t_{c,f}^{\rm start}$ and $t_{c,f}^{\rm end}$, the objective of beam service time allocation problem can be rewritten as
\begin{equation}
	\begin{aligned}
		\gamma_f & \triangleq \sum_{c=1}^C V\widetilde{D}_{c,f}-R_{c,f}Q_{c,f}(t_{c,f}^{\rm end}-t_{c,f}^{\rm start}+1) \\
        &\triangleq \sum_{c=1}^C (\frac{V}{f}+R_{c,f}Q_{c,f})t_{c,f}^{\rm start}-(R_{c,f}Q_{c,f}) t_{c,f}^{\rm end}.
	\end{aligned}
\label{new_expression}
\end{equation}

Since cells allocated with different frequency bands do not interfere with each other,
beam service time allocation problem can be simplified by just focusing on co-frequency cells.
Define auxiliary variables $m_{c,f}=V/f+R_{c,f}Q_{c,f}$ and $n_{c,f}=R_{c,f}Q_{c,f}$.
Then, the objective of beam service time allocation problem of co-frequency cells can be further expressed as follows.
\begin{equation}
	\begin{aligned}
		 \gamma_{b,f}&= \sum_{s=1}^S \sum_{c \in \mathcal{C}_{s,b,f}} (m_{c,f} t_{c,f}^{\rm start}-n_{c,f} t_{c,f}^{\rm end}) =\sum_{s=1}^S \gamma_{s,b,f},
	\end{aligned}
\label{new_expression}
\end{equation}
where $\mathcal{C}_{s,b,f}$ is the set of cells the satellite $s$ serves using a beam with the same spectrum as beam $b$ in scheduling epoch $f$.
Moreover, $\gamma_f = \sum_{b=1}^B \gamma_{b,f}/B$ and thus reducing $\gamma_{b,f}$ is equivalent to reducing $\gamma_f$.
In this case,
beam service time allocation problem on co-frequency cells can be formulated as follows:
\begin{align}
 	\boldsymbol{P_4}:	& \min\limits_{ \{t_{c,f}^{\rm start},t_{c,f}^{\rm end} |b, \forall s \}}  \quad	  \sum_{s=1}^S \gamma_{s,b,f} \label{STCA}\\
	s.t. \; \; &(\ref{p0}{a})-(\ref{p0}{c}), (\ref{eq:angle constraint2}), (\ref{gamma_s}). \notag
\end{align}

Assume cell indices in $\mathcal{C}_{s,b,f}$ are sorted according to the ascending order of initial service starting slot index, then
\begin{equation}
	\begin{aligned}
	&	\gamma_{s,b,f} =\sum_{c=1}^{|C_{s,b,f}|}  (m_{c,f} t_{c,f}^{\rm start}-n_{c,f}t_{c,f}^{\rm end})\\ & =(m_{1,f}t_{1,f}^{\rm start} - n_{|C_{s,b,f}|,f} t_{|C_{s,b,f}|,f}^{\rm end})\\ & + (m_{2,f} t_{2,f}^{\rm start}-n_{1,f}t_{1,f}^{\rm end})  \\&+...+ (m_{c,f}t_{c,f}^{\rm start}  -n_{c-1,f}t_{c-1,f}^{\rm end}) \\& +...+(m_{|C_{s,b}^f|,f} t_{|C_{s,b}^f|,f}^{\rm start}-n_{|C_{s,b}^f|-1,f} t_{|C_{s,b}^f|-1,f}^{\rm end}).
	\end{aligned}
	\label{gamma_s}
\end{equation}
where $|C_{s,b,f}|$ is the number of elements in set $\mathcal{C}_{s,b,f}$.
The beam service time allocation algorithm is summarized in Algorithm~\ref{alg:Service_Time_Allocation_Algorith}, which aims to minimize $\gamma_{s,b,f}$ by decreasing the value of each term $(m_{c,f}t_{c,f}^{\rm start}  -n_{c-1,f}t_{c-1,f}^{\rm end})$ in turn in (\ref{gamma_s}).
The cell visiting order follows the ascending order of initial beam service starting slot index and satellite index.
The procedure of Algorithm~\ref{alg:Service_Time_Allocation_Algorith} consists of 2 steps.
Firstly, it fixes the tuple $(t_{1,f}^{\rm start}, t_{|C_{s,b,f}|,f}^{\rm end})$ of all satellites and fine-tunes other tuples $(t_{c,f}^{\rm end}, t_{c+1,f}^{\rm start})$ in step 1 (line 3-9).
After all tuples in step 1 are adjusted, Algorithm~\ref{alg:Service_Time_Allocation_Algorith} then fine-tunes all tuple $(t_{1,f}^{\rm start}, t_{|C_{s,b,f}|,f}^{\rm end})$ in step 2 (line 10-14).
In each loop, Algorithm~\ref{alg:Service_Time_Allocation_Algorith} only fine-tunes a tuple of $(t_{c,f}^{\rm end}, t_{c+1,f}^{\rm start})$ considering the feasible beam service time allocation range while fixing the beam service time allocation decision of other cells.
Note that the interference constraints limit the feasible range of $t_{c,f}^{\rm start}$ and $t_{c,f}^{\rm end}$, which provide a lower bound $t_{c+1,f}^{\rm low}$ and an upper bound $t_{c,f}^{\rm up}$.
When $t_{c+1,f}^{\rm start} <t_{c+1,f}^{\rm low}$ or $t_{c,f}^{\rm end} > t_{c,f}^{\rm up}$, constraint (\ref{eq:angle constraint2}) is violated.
Moreover, $t_{c+1,f}^{\rm start}< D_{\rm max}-T+1+t_{c+1,f-1}^{\rm end}$ due to the maximum
beam revisit time constraint (\ref{p0}{a}).
Hence, $\min (t_{c+1,f}^{\rm end},D_{\rm max}-T+1+t_{c+1,f-1}^{\rm end}) > t_{c+1,f}^{\rm start} \geq \max(t_{c+1,f}^{\rm low}, t_{c,f}^{\rm start}+1)$ and $\min(t_{c,f}^{\rm up}, t_{c+1,f}^{\rm end}-1)> t_{c,f}^{\rm end} \geq t_{c,f}^{\rm start}$ (line 5-6).
Recalling that $\gamma_{s,b,f}$ is the sum of term $(m_{c+1,f}t_{c+1,f}^{\rm start}  -n_{c,f}t_{s,c,f}^{\rm end})$ and Algorithm~\ref{alg:Service_Time_Allocation_Algorith} reduces one term's value per loop while the other terms' values are fixed.
Therefore, the objective $\sum_{s=1}^S \gamma_{s,b,f}$ monotonically decreases after each beam service time adjustment, which ensures the convergence of Algorithm 2.

\begin{algorithm}[htbp]
	\begin{algorithmic}[1]
		\caption{Beam Service Time Allocation Algorithm}
		\label{alg:Service_Time_Allocation_Algorith}
		\footnotesize
		\State Input: $Q_{c,f}$, $\beta_{c,s,f}$, $b$, $T$, $\alpha_{c,b,f}$, $D_{\rm max}$, $t_{c,f-1}^{\rm end}$, $R_{c,f}$, initial beam service time allocation $(t_{c,f}^{\rm start}, t_{c,f}^{\rm end})$ provided by serving beam allocation algorithm.
		\State Construct the set $\mathcal{C}_{s,b,f}$ for all satellite and calculate $m_{c,f}$ and $n_{c,f}$ for cells in set $\mathcal{C}_{s,b,f}$.
		\For{$s=1:S$}
		\For{$c=1:|C_{s,b,f}|-1$}
		\State Calculate $t_{c,f}^{\rm up}$ and $t_{c+1,f}^{\rm low}$ based on constraint (\ref{eq:angle constraint2}).
		\State Find	the feasible ranges of $t_{c,f}^{\rm end}$ and $t_{c+1,f}^{\rm start}$.
		\State Obtain the optimal tuple ($t_{c,f}^{\rm end}, t_{c+1,f}^{\rm start})$ for minimizing $(m_{c+1,f}t_{c+1,f}^{\rm start}  -n_{c,f}t_{s,c,f}^{\rm end})$.
		\EndFor
		\EndFor
		\For{$s=1:S$}
		\State Calculate $t_{|C_{s,b,f}|,f}^{\rm up}$ and $t_{1,f}^{\rm low}$ based on (\ref{eq:angle constraint2}).
		\State Find	the feasible ranges of $t_{|C_{s,b,f}|,f}^{\rm end}$ and $t_{1,f}^{\rm start}$.
		\State Obtain the optimal tuple $(t_{|C_{s,b,f}|,f}^{\rm end},t_{1,f}^{\rm start})$ for minimizing $(m_{1,f}t_{1,f}^{\rm start} - n_{|C_{s,b,f}|,f} t_{|C_{s,b,f}|,f}^{\rm end})$.
		\EndFor
		\State Output:  $(t_{c,f}^{\rm start},t_{c,f}^{\rm end})$ for cells in set $\mathcal{C}_{s,b,f}$.
	\end{algorithmic}
\end{algorithm}

\subsection{Serving satellite allocation problem and algorithm design}\label{sec:subproblem 3}

In LEO satellite networks, given the dynamics of network topology, it is essential to update the service relationship between satellites and cells.
Note that serving satellite allocation $\beta_{c,s,f}$ tightly couples with serving beam allocation and beam service time allocation, and their closed-form relationships cannot be derived.
Hence, we propose a meta-heuristic serving satellite allocation algorithm shown in Algorithm~\ref{alg:service_satellite_allocation_Algorith}, which is based on simulated annealing procedure. At the gateway station, the algorithm is implemented periodically or when a serving satellite cannot provide service to a cell, i.e., the elevation angle is below the minimum.

The main steps of Algorithm~\ref{alg:service_satellite_allocation_Algorith} is as follows.
Firstly, serving satellites for cells are initially determined based on the serving satellite allocation of the last scheduling epoch (line 3).
Subsequently, initial service duration $t_{c,f}$ is calculated for each cell, which is an input of serving beam allocation algorithm (line 4).
By invoking Algorithm~\ref{alg:BA_Algorithm} and Algorithm~\ref{alg:Service_Time_Allocation_Algorith}, the performance of current satellite-cell service relationship is evaluated   and it is regarded as the current best allocation result (line 5-6).
In following loops, new serving satellite allocations are generated based on the solution search approach of simulated annealing procedure.
If the performance of new satellite allocation is better than the current one, set it as the current satellite allocation (line 12-15).
Otherwise, set it as the current satellite allocation with pre-defined probability (line 18-19).
Finally, Algorithm~\ref{alg:service_satellite_allocation_Algorith} outputs the solution $\{(\alpha_{c,b,f}, \beta_{c,s,f},  t_{c,f}^{\rm start}, t_{c,f}^{\rm end})\}$ of problem $\boldsymbol{P_2}$  (line 23).	

\begin{algorithm}[htbp]
\begin{algorithmic}[1]
	\caption{Serving Satellite Allocation Algorithm}
	\label{alg:service_satellite_allocation_Algorith}
    \footnotesize
	\State Input: $Q_{c,f}$, $T$, $\beta_{c,s,f-1}$, $\widetilde{D}_{c,f-1}$, $t_{c,f-1}^{\rm end}$, $B$, $R_{c,f}$.
	\State Initialization:  initial temperature $T_1$, final temperature $T_2$, cooling schedule $\varrho\in (0,1)$, the best satellite allocation $\{\beta_{c,s,f}\}$, current satellite allocation $\{{\beta_{c,s,f}^\prime}\}$,  best objective value $\gamma_f^{best}$, queue length of satellite $Q_{s,f}$.
	\State Generate a feasible set $\{\beta_{c,s,f}\}$ based on $\{\beta_{c,s,f-1}\}$ and calculate $Q_{s,f}$: $Q_{s,f}=\sum_{c=1}^C {\beta_{c,s,f}}Q_{c,f} $.
	\State Calculate the service duration $t_{c,f}$ for each cell: $t_{c,f} \gets \min ( \frac{Q_{c,f}TB }{Q_{s,f}}, \lceil \frac{Q_{c,f}}{R_{c,f}} \rceil)$.
	\State Invoke  \textbf{Algorithm~\ref{alg:BA_Algorithm}} and \textbf{Algorithm~\ref{alg:Service_Time_Allocation_Algorith}} based on $\{\beta_{c,s,f}\}$ and $\{t_{c,f} \}$ and calculate $\gamma_f $.
	\State $\{\beta_{c,s,f}^\prime \} \gets \{\beta_{c,s,f}\}$, $\gamma_f^{best} \gets \gamma_f$.
	\While{$T_1>T_2$ }
	\State $T_1 \gets \varrho T_1$.
	\State Randomly generate a new $\{\beta_{c,s,f}^{\prime \prime}\}$  based on $\{\beta_{c,s,f}^\prime\}$ and calculate the service duration $t_{c,f} $ for each cell.
	\State Invoke  \textbf{Algorithm~\ref{alg:BA_Algorithm}} and \textbf{Algorithm~\ref{alg:Service_Time_Allocation_Algorith}} obtain a new $\gamma_f^\prime $.
	\If{$\gamma_f - \gamma_f^\prime>0$}
	\State $\{\beta_{c,s,f}^\prime\} \gets \{\beta_{c,s,f}^{\prime \prime}\}$, $\gamma_f \gets \gamma_f^\prime $.
	\If{ $\gamma_f^\prime >\gamma_f^{best}$}
	\State $\{\beta_{c,s,f}\} \gets \{\beta_{c,s,f}^{\prime \prime}\}$, $\gamma_{best} \gets \gamma_f^\prime $.
	\EndIf
	\Else
	\If{randomly value $\omega \geq exp(-(\gamma_f - \gamma_f^\prime)/T_1)$}
	\State $\{\beta_{c,s,f}^\prime\} \gets \{\beta_{c,s,f}^{\prime \prime}\}$, $\gamma_f \gets \gamma_f^\prime $.
	\EndIf
	\EndIf
	\EndWhile
	\State Output: $\{(\alpha_{c,b,f}, \beta_{c,s,f},  t_{c,f}^{\rm start}, t_{c,f}^{\rm end})\}$ for all cells.
	\end{algorithmic}
\end{algorithm}

\subsection{Convergence and complexity analysis}\label{sec:Complexity Alalysis}
\subsubsection{Convergence analysis}

Note that our proposed beam resource management approach is actually Algorithm 3, which incorporates Algorithm 1 and 2.
For the outer loop in Algorithm 3, it follows simulated annealing procedure, whose convergence is guaranteed via gradually decreasing temperature parameter.
Meanwhile, Algorithm 1 only executes finite steps and the convergence of Algorithm 2 has been proved in subsection B under any given serving satellite allocation among cells.
Therefore, Algorithm 3 is ensured to converge.

\subsubsection{Complexity analysis}For serving beam allocation algorithm, its complexity mainly relies on visiting all vertices to search the maximum weighted independent set.
Note that our proposal requires visiting all vertices and edges, and hence the complexity of the proposed algorithm is $\mathcal{O}(|\mathcal{V}|+|\mathcal{E}|)$, where $|\mathcal{V}|$ and $|\mathcal{E}|$ represent the number of vertices and edges, respectively.
Denote the maximum beam number of a satellite as $B_s^{\rm max}$.
Recalling that the number of cells and time slots in a scheduling epoch are $C$ and $T$, the maximum values of $|\mathcal{V}|$ and $|\mathcal{E}|$ are $CTB_s^{\rm max}$ and $\frac{CTB_s^{\rm max}(CTB_s^{\rm max}-1)}{2}$, respectively.
Hence, the complexity of serving beam allocation algorithm is further expressed as $\mathcal{O}((CTB_s^{\rm max})^2)$.
The beam service time allocation algorithm performs $C$ fine-tuning procedures for $C$ cells.
For the current visited cell, we need to check the feasible time allocation range of $t_{c,f}^{\rm start}$ and $t_{c,f}^{\rm end}$ for most $2S_{\rm max}T$ times, where $S_{\rm max}$ represents the maximum number of visible satellites with $S_{\rm max} <C$ and the ranges of $t_{c,f}^{\rm start}$ and $t_{c,f}^{\rm end}$ is smaller than $T$.
Therefore, the complexity of beam service time allocation algorithm is $\mathcal{O}(CT^2S_{\rm max})$.
The complexity for updating parameters of simulated annealing procedure is  $\mathcal{O}(J)$ with $J $ being the pre-set maximum iteration number.
For the serving satellite allocation algorithm, it invokes serving beam allocation and beam service time allocation algorithms in each iteration and the complexity of it is $\mathcal{O}(J(CTB_s^{\rm max})^2)$.
Finally, the complexity to obtain the solution of beam management $\{(\alpha_{c,b,f}, \beta_{c,s,f},  t_{c,f}^{\rm start}, t_{c,f}^{\rm end})\}$ to problem $\boldsymbol{P_2}$
is $\mathcal{O}(J(CTB_s^{\rm max})^2)$.

\section{Simulation Results and Analysis}\label{sec:sim}

In this section, extensive simulations are conducted to evaluate the performance of the proposed beam management approach in LEO satellite networks, in terms of average beam revisit time, cell average queue length, inter-satellite handover frequency, and the objective value of problem $\boldsymbol{P_0}$.
Specifically, we first introduce simulation parameters and verify the superiority of our proposed serving beam allocation and beam service time allocation algorithms.
Then, we investigate and analyze the impacts of dynamic LEO satellite topologies and time-varying traffic on network performance.

\subsection{Simulation parameters, comparison baselines and performance metrics}

\begin{table}[htbp]
\centering
\caption{Simulation parameter setting}
\label{tab:1}
\begin{tabular}{llllll} \toprule
\textbf{Parameter}                                     & \textbf{Value} \\
\midrule
The number of satellite orbits            & 40   \\
The center location of the concerned area & $(110.6^{\circ}E,26.67^{\circ}N)$ \\
 The number beams per satellite            & 4     \\
The number of satellites in an orbit      & 30   \\
The number of cells in the concerned area                     & 42         \\
 Frequency reuse factor                    & 4     \\
Orbit altitude                            & 600 km  \\
The number of slots in a scheduling epoch & 15     \\
$INR^{\rm th}$       & -10 dB      \\
Orbit inclination               & $50^{\circ}$   \\
Downlink target SNR of cells              & 20 dB   \\
$G_{\rm max}/G_{\rm min}$        &  30 dB   \\
The radius of a cell                 & 43.3 km  \\
Beam bandwidth                    & 500 MHz    \\
$G^{\rm th}$                  & -51 dB \\
$D_{\rm max}$                 & 50 slots  \\
Center oprtation frequency                    & 30 GHz    \\
Minimum elevation angle                   & $40^{\circ}$\\
\bottomrule
\end{tabular}
\end{table}

An Walker constellation with 1200 satellites evenly distributed across 40 orbits is constructed by AGI Systems Tool Kit (STK), where the inclination and altitude of satellite orbits are $50^{\circ}$ and 600 km, respectively.
Each satellite operates at Ka band with the downlink center frequency being 30 GHz, and 2 GHz spectrum is shared by all satellites.
Each satellite is with 4 beams that equally share 2 GHz bandwidth and that is the bandwidth of each beam is 500 MHz.
The downlink target SNR $SNR_{c}$ of all cells is 20 dB to simplify simulations, and the pre-fixed INR threshold $INR^{\rm th}$  is -10 dB.
In addition, 42 earth-fixed cells with hexagonal shapes arranged in 6 rows and 7 columns are considered, and the distance between the centers of two adjacent cells is 50 km.
The average data arrival rates of cells are generated by referring to \cite{cell_traffic}.
There are 15 time slots per scheduling epoch and the maximum  beam revisit time of cells is 50 time slots.

The gain ratio of $G_{\rm max}/G_{\rm min}$  is set as 30 dB.
The center location of the concerned area is  $(110.6^{\circ}E,26.67^{\circ}N)$.
In addition, rain attenuation is the main factor causing the pathloss variation, which depends on the rainfall rate.
Here, the minimum elevation angle of UEs is $40^{\circ}$ and rain attenuation is calculated based on point rainfall rate for the location for 1$\%$ of an average year.
According to ITU-R model\cite{ITU}, when satellites operate at 30 GHz, the attenuation caused by rain, atmospheric gases, cloud, and fog has 99$\%$ probability of less than 10 dB in Beijing and 15 dB in Sanya.
Note that the latitude of the concerned area is between the latitudes of Beijing and Sanya, and the average rainfall rate generally decreases with increasing latitude in China.
Based on the above analysis, we set $G^{\rm th}$   in (\ref{eq:angle_constraint1}) as -51 dB.
In addition, if accurate meteorological data information can be obtained, $G^{\rm th}$  can be dynamically adjusted.

Main parameters are summarized in Table~\ref{tab:1}.
The simulation is done with MATLAB.
Three benchmark schemes are compared to our proposed serving beam allocation and beam service time allocation algorithms that constitute the frequency-time resource management scheme.
\begin{itemize}
	\item \emph{Greedy scheme:} The gateway station allocates serving beam and service time to cells based on greedy algorithm \cite{greedy2}.
	\item \emph{GA based scheme:} The gateway station allocates serving beam and service time to cells by genetic algorithm \cite{ga1}, and the algorithm is periodically executed in each slot. To limit the solution searching space, the resource configuration of one satellite is optimized each time.
     \item \emph{Swap based scheme:} The gateway station allocates serving beam and service time to cells based on swap matching algorithm in \cite{swap}. In this scheme, the initial beam management plan is obtained by greed scheme, and then the serving times and beams of two cells will be continuously swapped if a swap operation can reduce the objective function value of problem $\boldsymbol{P_2}$.
\end{itemize}
Meanwhile, three baselines are considered for serving satellite allocation optimization.
\begin{itemize}
	\item \emph{Minload:} Each cell is allocated with the satellite with the minimum load \cite{traditional_handover_1}.
	\item \emph{Maxtime:} Each cell is allocated with the satellite that provides the maximum service duration \cite{traditional_handover_3}.
	\item \emph{TOPSIS:} A multi-attribute decision based scheme adopted by \cite{multi-attribute_1}.
\end{itemize}
As for performance metrics, we take the average among all cells. Hence, performance metrics include average beam revisit time, average queue length, average service duration length, and average handover number.

\subsection{The performance of the serving beam allocation and beam service time allocation algorithms}

\begin{figure}[htbp]
\centering
\subfigure[]
{
\includegraphics[width=3in]{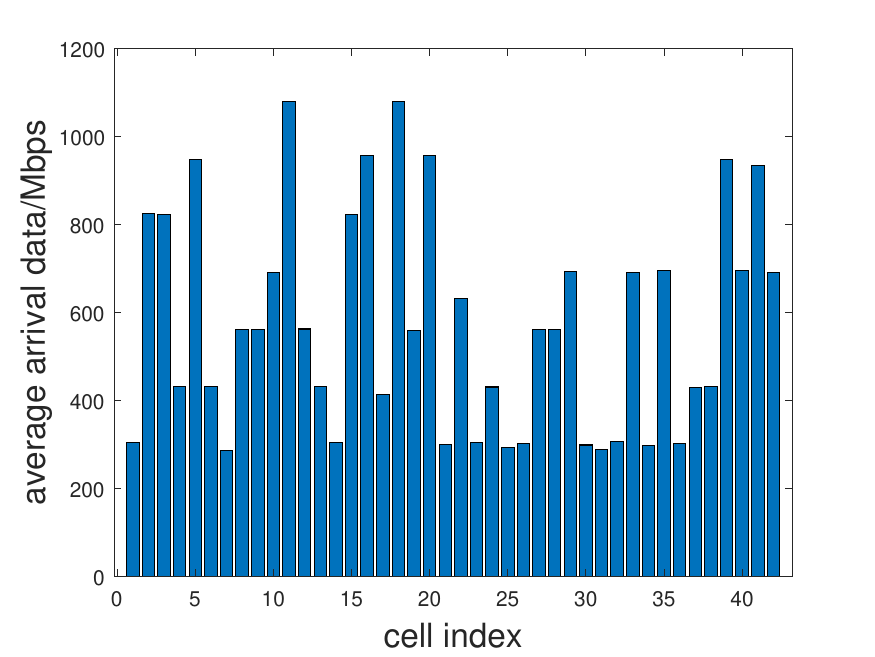}
}
\subfigure[]
{
\includegraphics[width=3in]{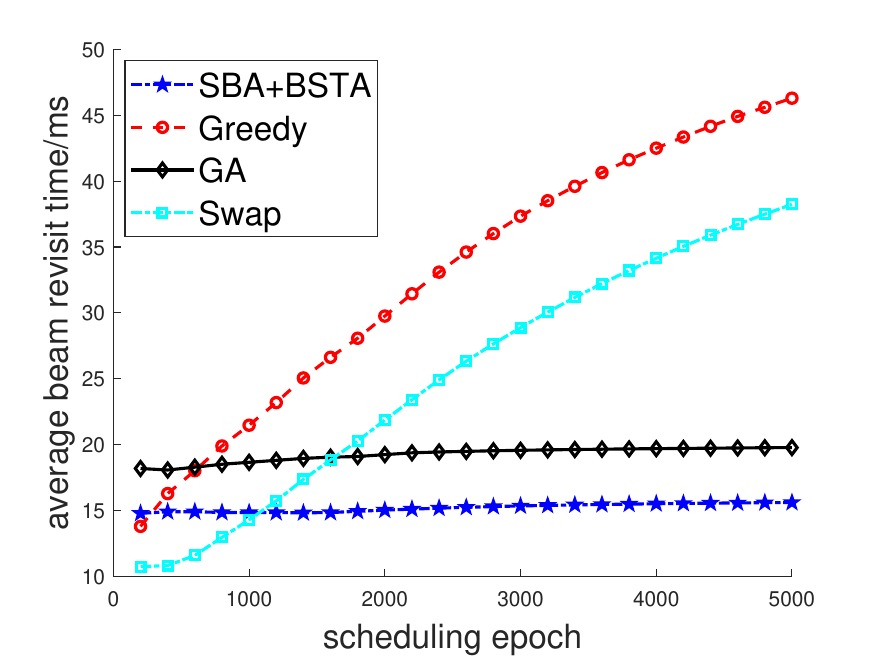}
}
\caption{Figure (a) is the unbalanced cell traffic demand expectation in simulations and (b) shows the average beam revisit time of cells.}\label{fig5}
\end{figure}
In this part, the duration of each scheduling epoch is set to 20 ms.
The serving beam allocation and beam service time allocation algorithms are denoted as \textquotedblleft SBA\textquotedblright~and \textquotedblleft BSTA\textquotedblright, respectively.
The generated cell traffic demand expectation is illustrated in Fig.~\ref{fig5}(a).
Fig.~\ref{fig5}(b) shows the average beam revisit time under different time-frequency resource management schemes with the control parameter $V=1000$, while they adopt the same multi-attribute decision based satellite allocation scheme during the simulation for fair comparisons.
It can be seen that the beam revisit time of our proposal is about 14.8 - 15.8 ms, which reduces 20.8$\%$, $59.12\%$, and 65.98$\%$ compared with genetic algorithm based scheme, swap based scheme,  and greedy algorithm based scheme, respectively.
In addition, the beam revisit time of a cell with baseline schemes can be longer than the duration of a scheduling epoch and this is because interference constraints prevent some cells from being served within a scheduling epoch.
Moreover, it can be observed that the average beam revisit time with a greedy algorithm-based and swap schemes are significantly larger than others.
Meanwhile, they achieve shorter beam revisit time at the beginning of simulations, and
the reason is concluded as follows.
In the first scheduling epoch, the initial beam revisit time of cell $c$ is equal to $t_{c,1}^{\rm start}$, which makes the greedy algorithm and swap matching algorithm allocate time-frequency resource to cells with less queue length.
Therefore, they obtain a small beam revisit time at the beginning of simulations.
However, with the scheduling epoch increasing, the effect of cell queue length becomes more significant according to (\ref{eq:gamma}).
At this time, the greedy and swap matching algorithms prefer to allocate longer service duration to cells with larger queue length, making more cells unable to obtain service due to interference constraints.

\begin{figure}[htbp]
    \centering
        \subfigure[]
		{
                \includegraphics[width=2.8in]{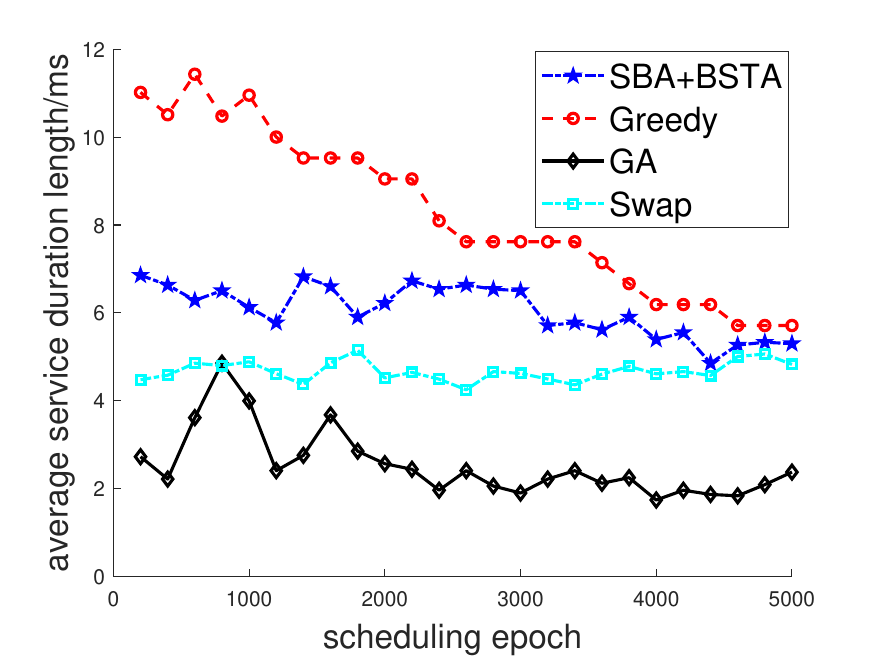}
        }
        \subfigure[]
		{
                \includegraphics[width=2.8in]{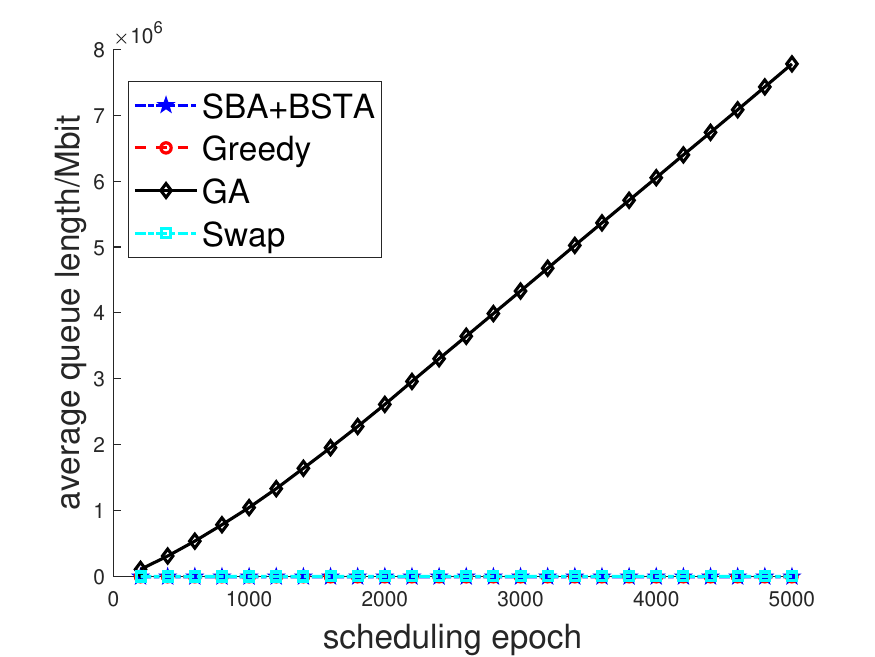}
        }
        \subfigure[]
		{
                \includegraphics[width=2.8in]{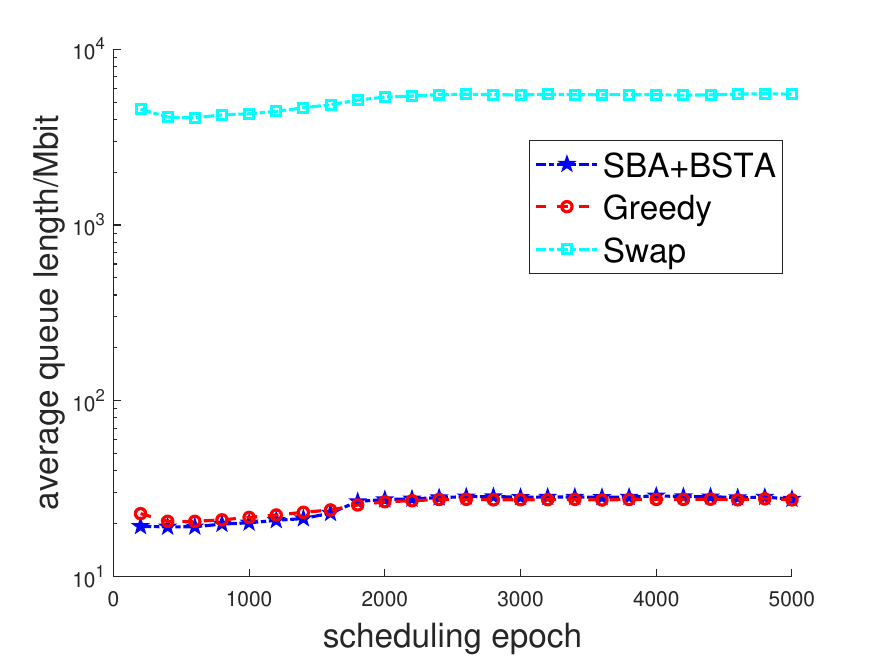}
        }
    \caption{Average beam service duration length and average queue length comparison among different frequency-time resource management schemes.}\label{fig6}
\end{figure}

In Fig.~\ref{fig6}(a), it can be seen that genetic algorithm based scheme leads to the shortest beam service time length of cells, making the network unstable as shown in Fig.~\ref{fig6}(b). This is because
genetic algorithm based scheme frequently changes served cells in two adjacent slots due to the randomness of solution update.
However, such randomness also contributes to lower beam revisit time as shown in Fig.~\ref{fig5}(b).
In addition, since the average queue length curves of our proposal, greedy algorithm,  and swap matching algorithm are close in Fig.~\ref{fig6}(b), we independently amplify them in Fig.~\ref{fig6}(c).
Moreover, as shown in Fig.~\ref{fig6} (a)-(c), our proposed serving beam allocation and beam service time allocation algorithms achieve similar
beam service time length and queue length compared to greedy algorithm based scheme. Nevertheless, greedy algorithm based scheme
leads to higher beam revisit time\ as shown in Fig.~\ref{fig5}(b).
Although swap matching algorithm can reduce the beam revisit time compared with greedy algorithm, the average service duration length is less than greedy algorithm, which causes longer queue length.

\begin{figure}[htbp]
    \centering

        \subfigure[]
		{
                \includegraphics[width=2.8in]{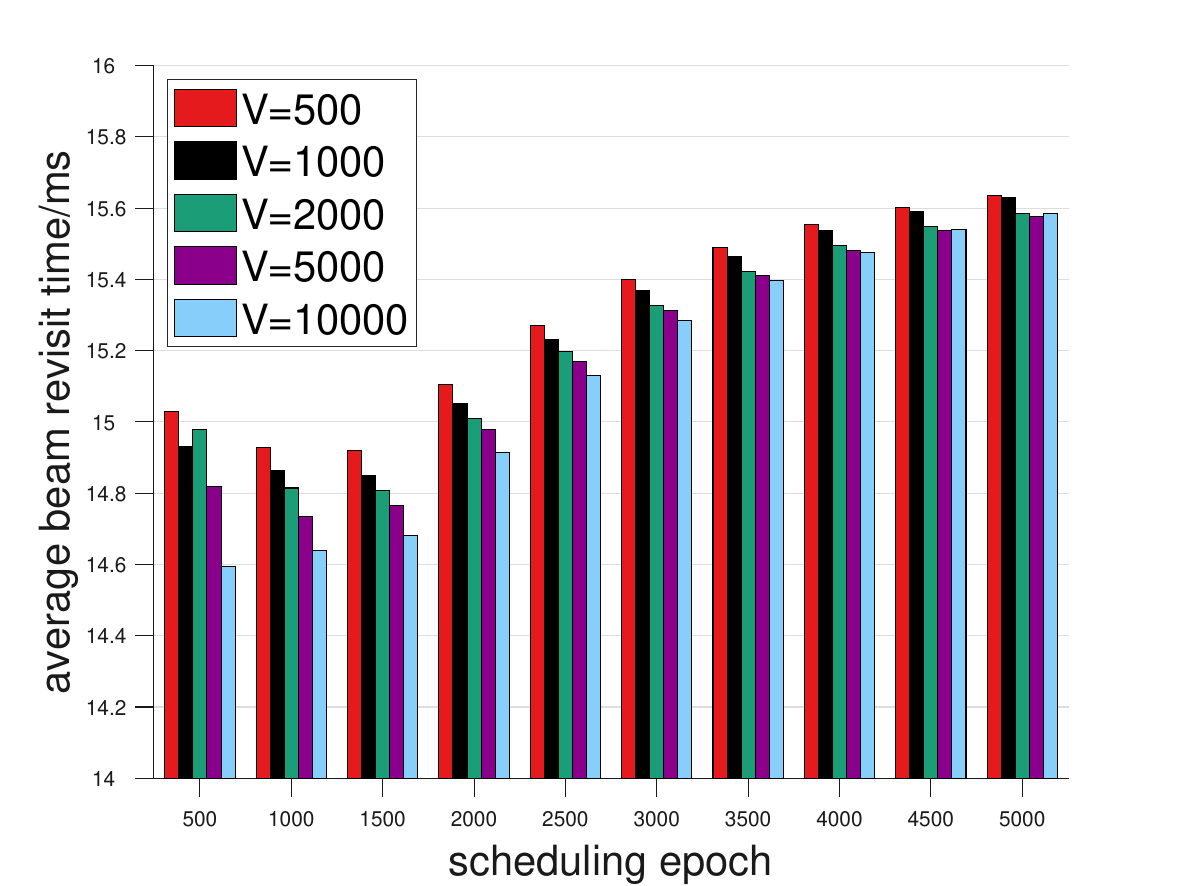}
        }
        \subfigure[]
		{
                \includegraphics[width=2.8in]{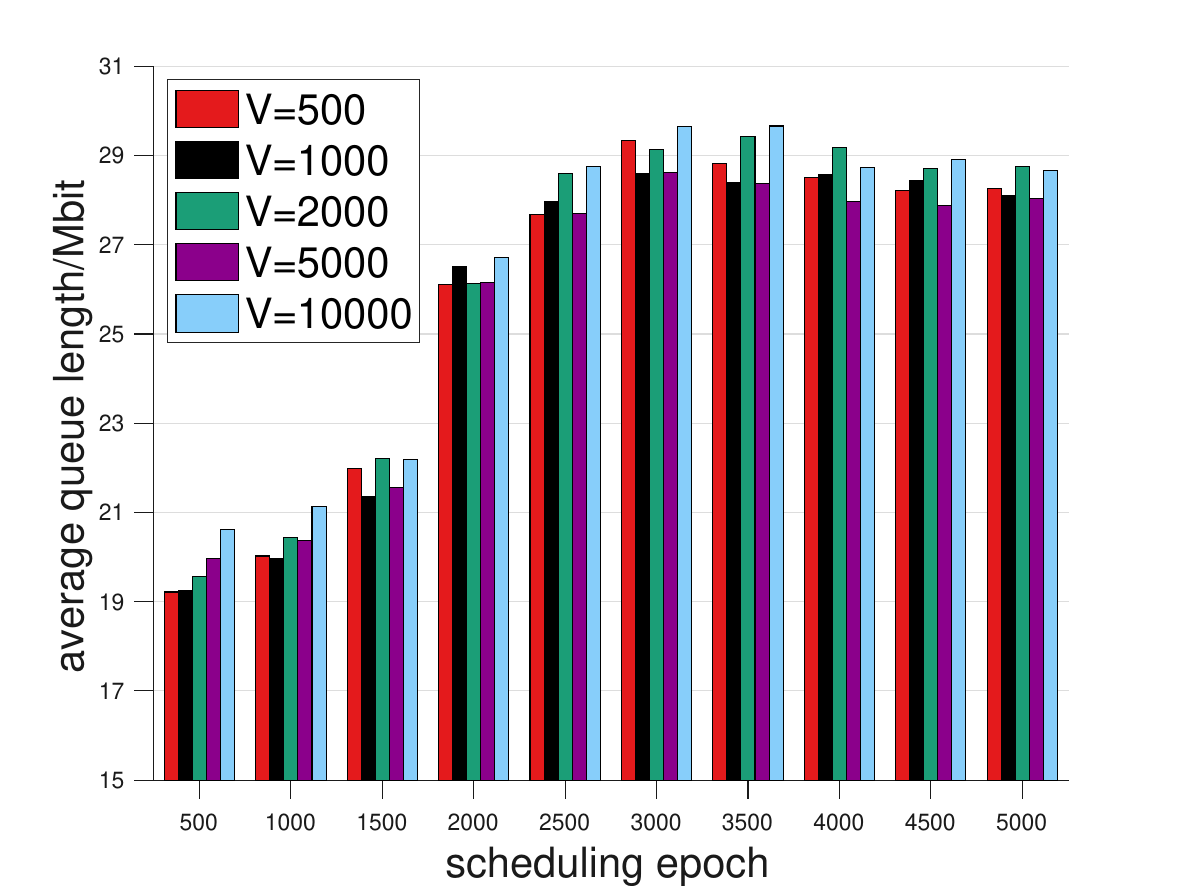}
        }
    \caption{The influence of control parameter $V$ on average beam revisit time and average queue length of cells.}\label{fig7}
\end{figure}

\begin{figure}[htbp]
    \centering

        \subfigure[]
		{
                \includegraphics[width=1.625in]{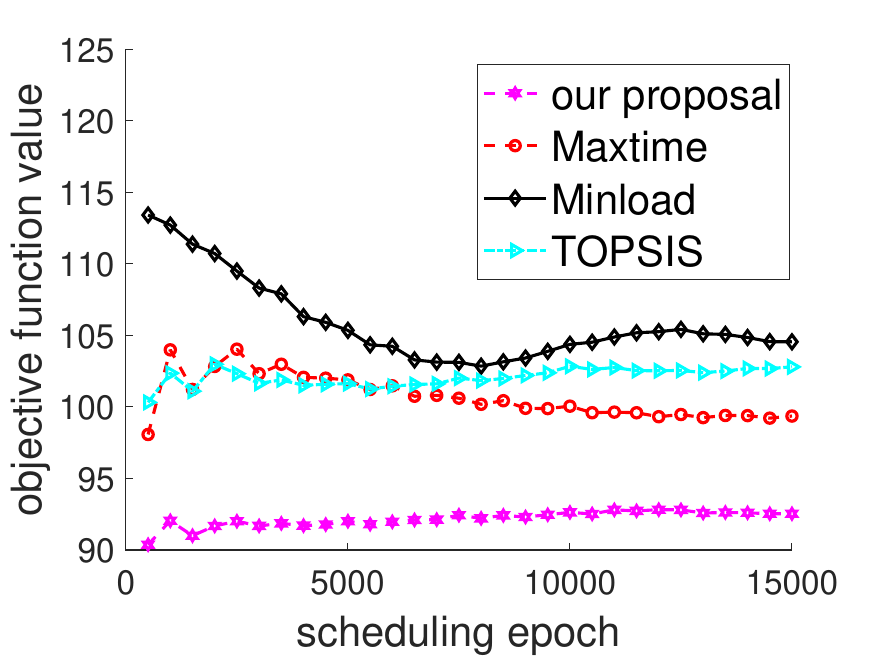}
        }
        \subfigure[]
		{
                \includegraphics[width=1.625in]{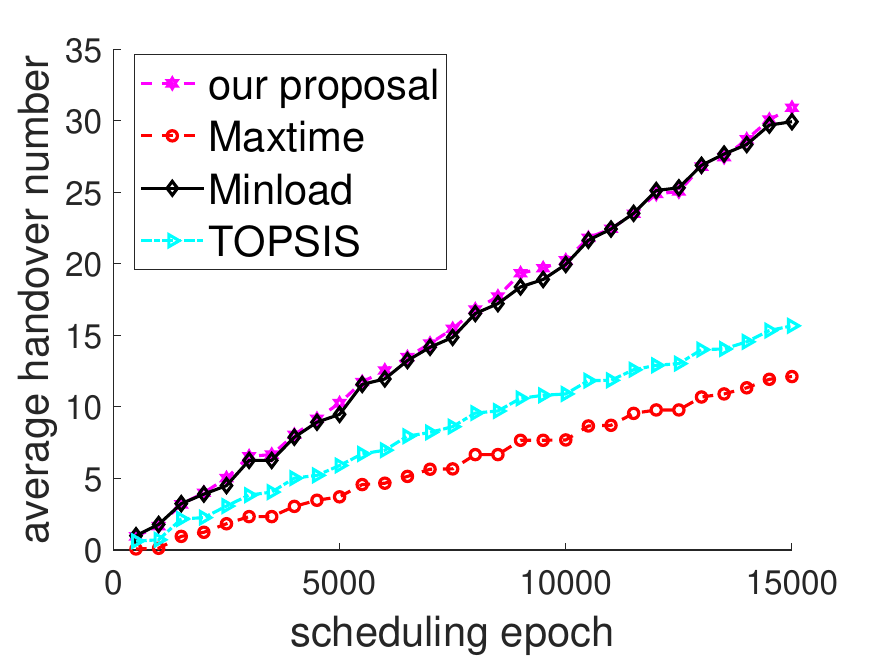}
        }
        \subfigure[]
		{
                \includegraphics[width=1.625in]{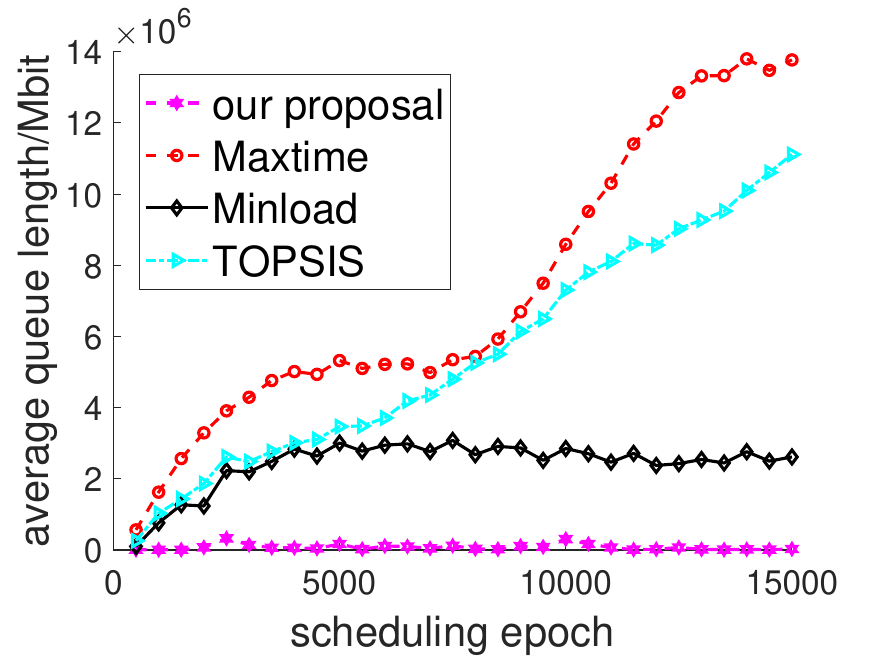}
        }
        \subfigure[]
		{
                \includegraphics[width=1.625in]{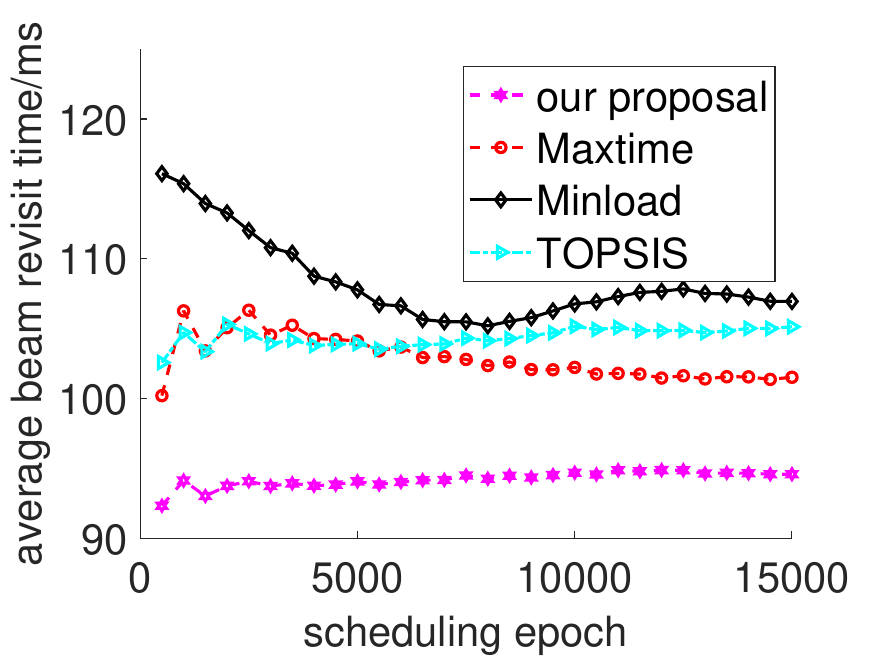}
        }
    \caption{The performance comparison among our proposal and baselines in the concerned dynamic LEO satellite network.}\label{fig8}
\end{figure}

Fig.~\ref{fig7} shows the influence of control parameter $V$ on average beam revisit time and average queue length of cells.
According to the result, it can be observed that choosing a large $V$ decreases the average beam revisit time of cells.
Moreover, the average cell beam revisit time fluctuates under different $V$, which is led by the change in the total number of visible satellites of cells due to satellite movement.
On the other hand, although it is shown in~\cite{Queueing_object} that a smaller $V$ contributes to a lower average queue length,
in our simulation, changing $V$ cannot significantly affect the average queue length as shown in Fig.~\ref{fig7}(b).
This is because the average queue length in our case not only depends on the beam service duration of each cell but also is influenced by interference constraints, serving beam allocation, and the number of visible satellites.

\begin{figure}[htbp]
    \centering

        \subfigure[]
		{
                \includegraphics[width=2.8in]{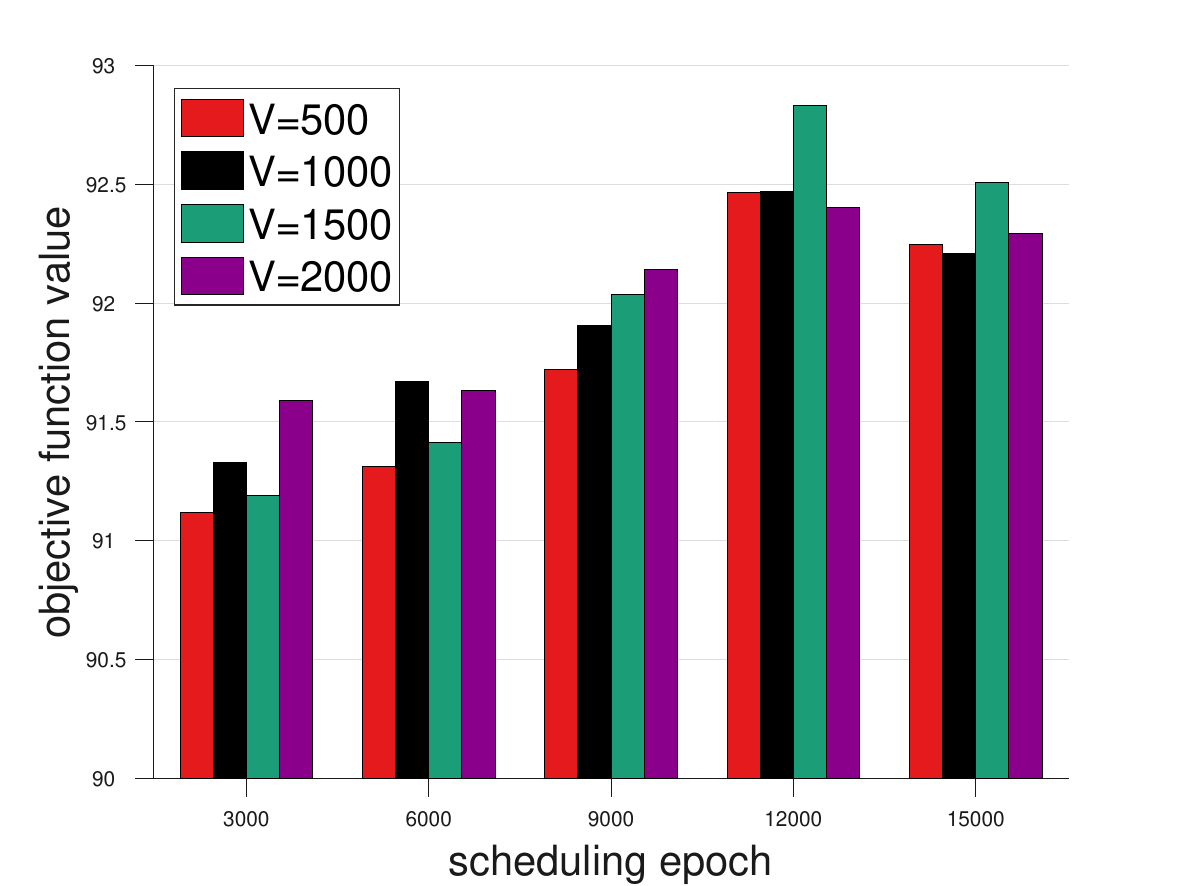}
        }
        \subfigure[]
		{
                \includegraphics[width=2.8in]{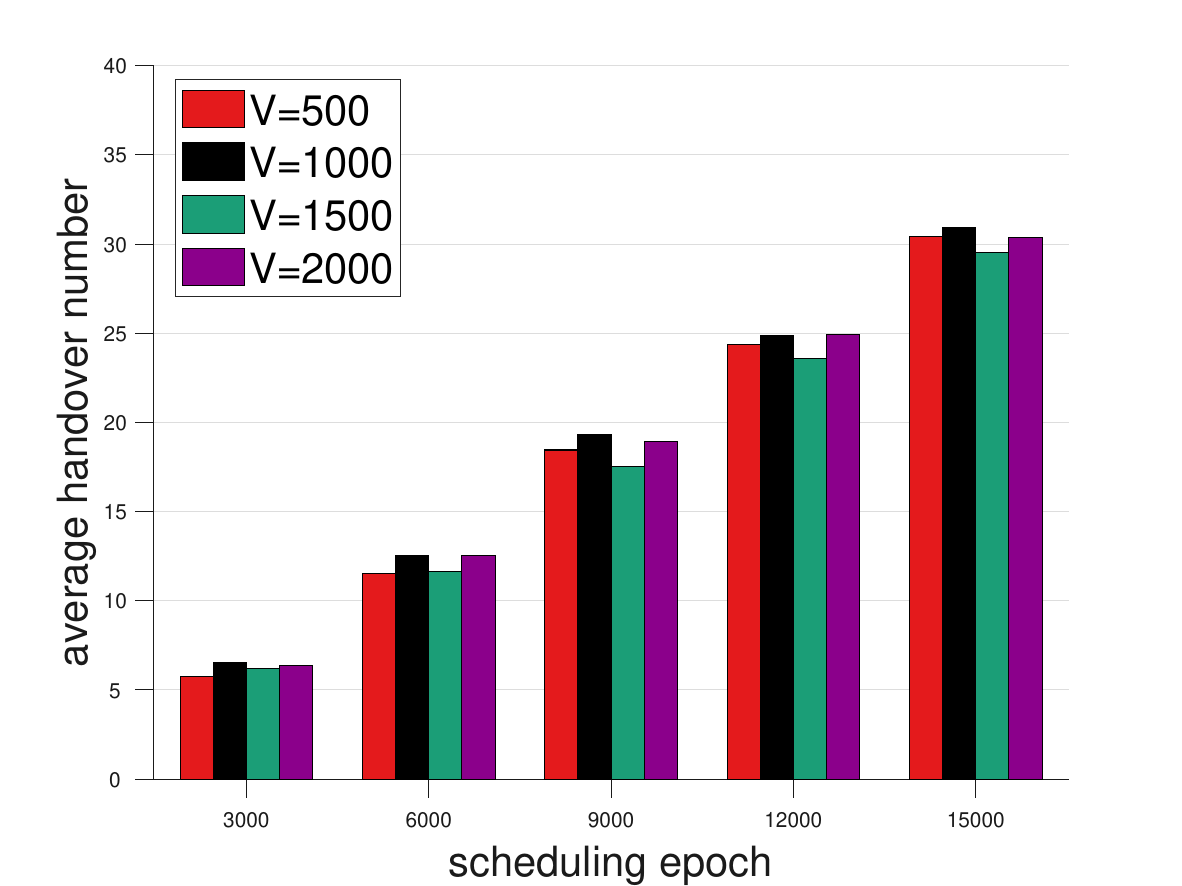}
        }
        \subfigure[]
		{
                \includegraphics[width=2.8in]{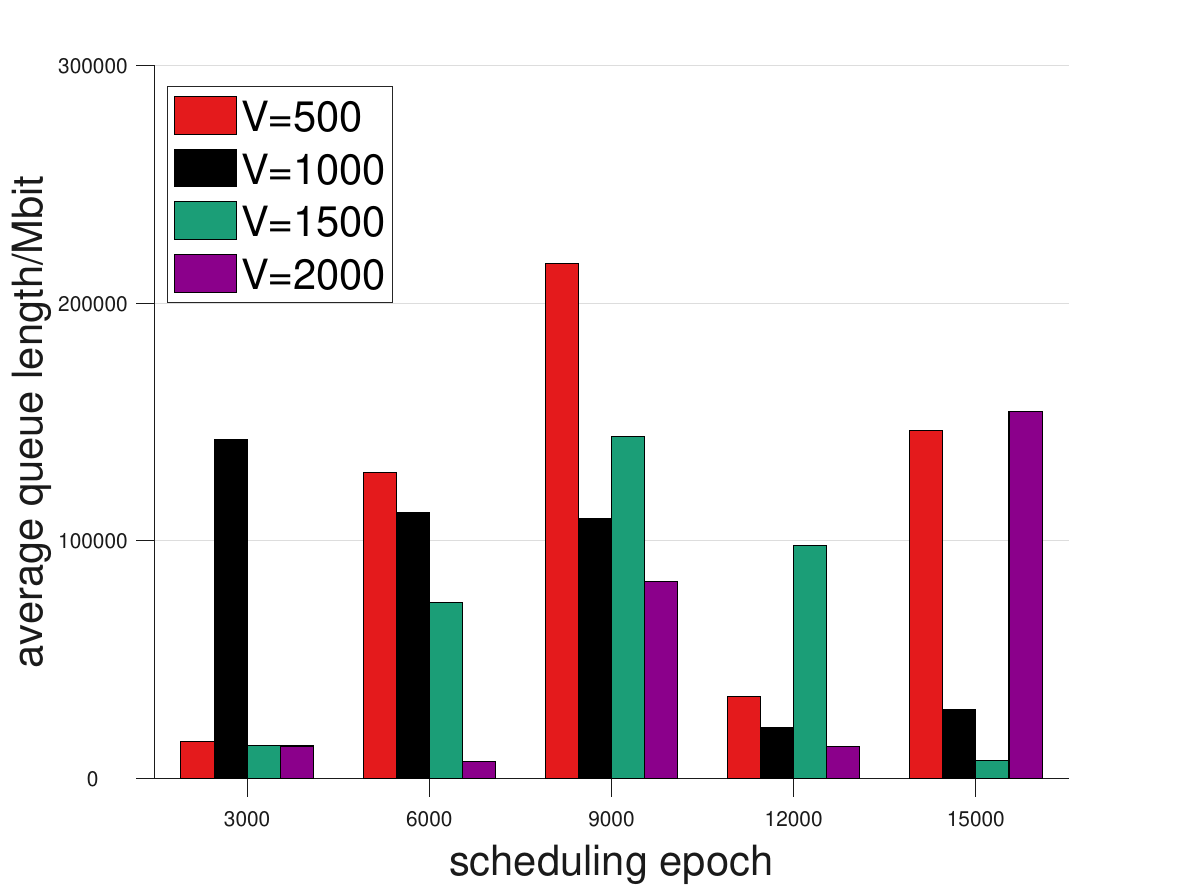}
        }
        \subfigure[]
		{
                \includegraphics[width=2.8in]{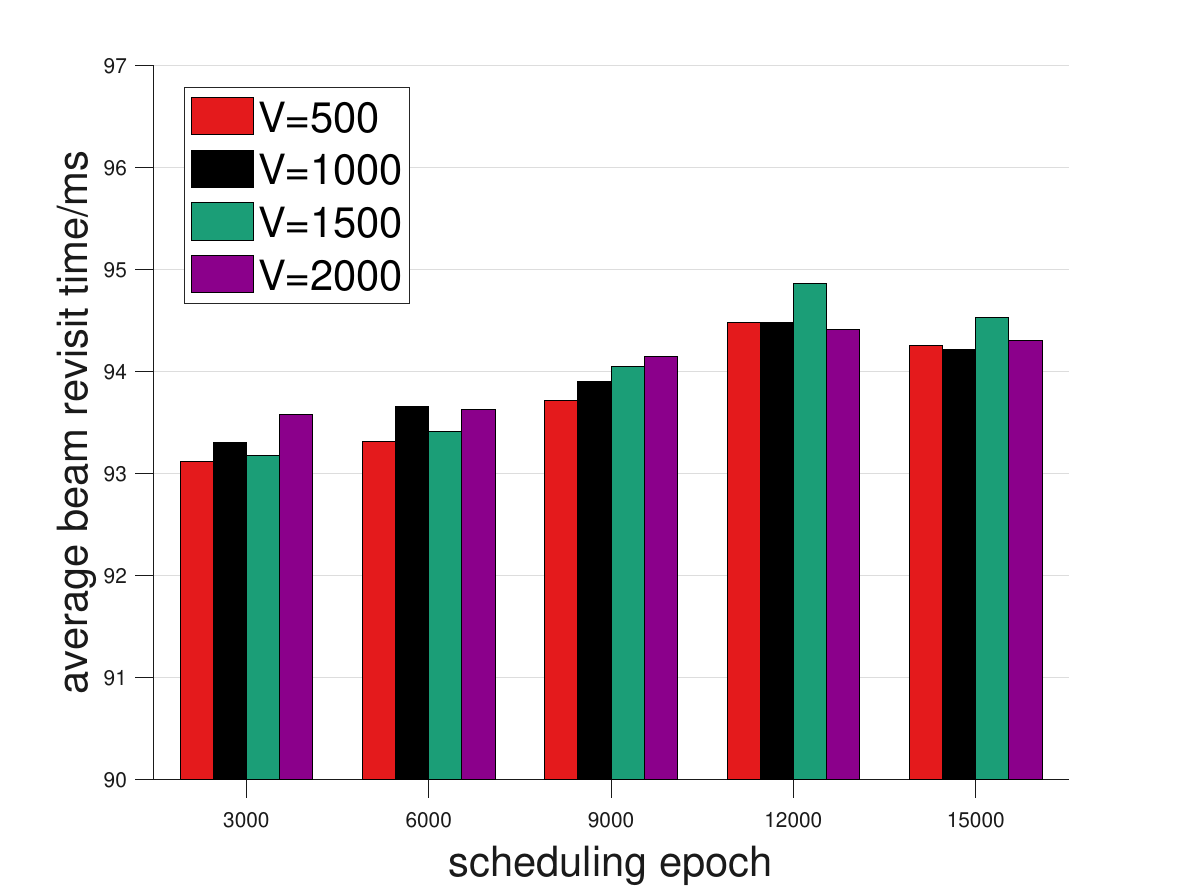}
        }
    \caption{The performance of the proposed serving satellite allocation algorithm under different V in the concerned dynamic LEO satellite network.}\label{fig12}
\end{figure}

\subsection{The performance of the proposed serving satellite allocation algorithm}

To better observe the influence of time-varying network topology, we increase the number of scheduling epochs and extend the duration of each scheduling epoch to reduce the simulation data size and complexity.
In this subsection, the scheduling epoch is set to 120 ms with 15 slots and the simulation duration is set to 30 minutes.
The proposed serving satellite allocation algorithm and baselines are executed every 600 scheduling epochs
or when a serving satellite can no longer provide service to a cell.
Meanwhile, the baselines adopt Algorithm 1 and 2 for serving beam allocation and beam service time allocation to make fair comparisons.
Fig.~\ref{fig8} illustrates the performance comparison result of our proposal with benchmark schemes in the concerned dynamic LEO satellite network.
As shown in Fig.~\ref{fig8} (a), our proposal achieves the minimal average beam revisit time, which reduces 7.2$\%$, 10.39$\%$ and 11.91$\%$ compared with $Maxtime$ method, $TOPSIS$ method and $Minload$ method when the scheduling epoch index reaches 15000.
Moreover, our proposal also results in the minimum average queue length as illustrated in Fig.~\ref{fig8} (b).
Fig.~\ref{fig8} (c) indicates that our proposal achieves a similar inter-satellite handover frequency with $Minload$ method, and the average handover interval is 501 scheduling epochs, i.e., around 1 minute.
In addition, the objective values of problem $\boldsymbol{P_0}$ are shown in Fig.~\ref{fig8} (d).
Our proposal outperforms all the baselines because it adjusts the serving relationships among cells and satellites from a global perspective and achieves better beam revisit time as well as comparable inter-satellite handover frequency.

Next, we examine the influence of the control parameter $V$, where the proposed serving satellite allocation algorithm is executed every 600 scheduling epochs.
According to Fig.~\ref{fig12} (a), a different conclusion is made compared with Fig.~\ref{fig7} (a).
Specifically, although we set a large $V$ to make algorithms pay more attention to reducing  beam revisit time, but the average  beam revisit time may still increase.
The reason is that numerous factors affect the state of networks and algorithm performance, including dynamic network topologies, the number of available serving satellites, current and past serving relationships between satellites and cells, dynamic inter-cell interference and the arrival of data packets.
In addition, as shown in Fig.~\ref{fig12} (a)-(d), the average beam revisit time, the number of inter-satellite handover, average queue length, and the objective value of problem $\boldsymbol{P_0}$ slightly jitter when setting different values of control parameter $V$.
Nevertheless, it is observed that our proposal performs better than baselines as shown in Fig.~\ref{fig8}.

\begin{figure}[htbp]
    \centering
        \subfigure[]
		{
                \includegraphics[width=2.8in]{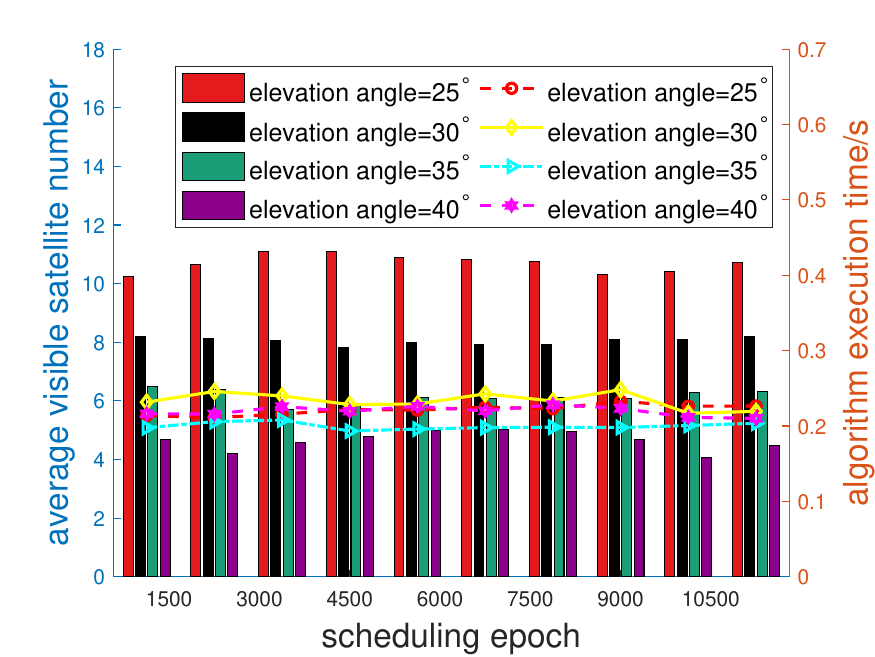}
        }
        \subfigure[]
		{
                \includegraphics[width=2.8in]{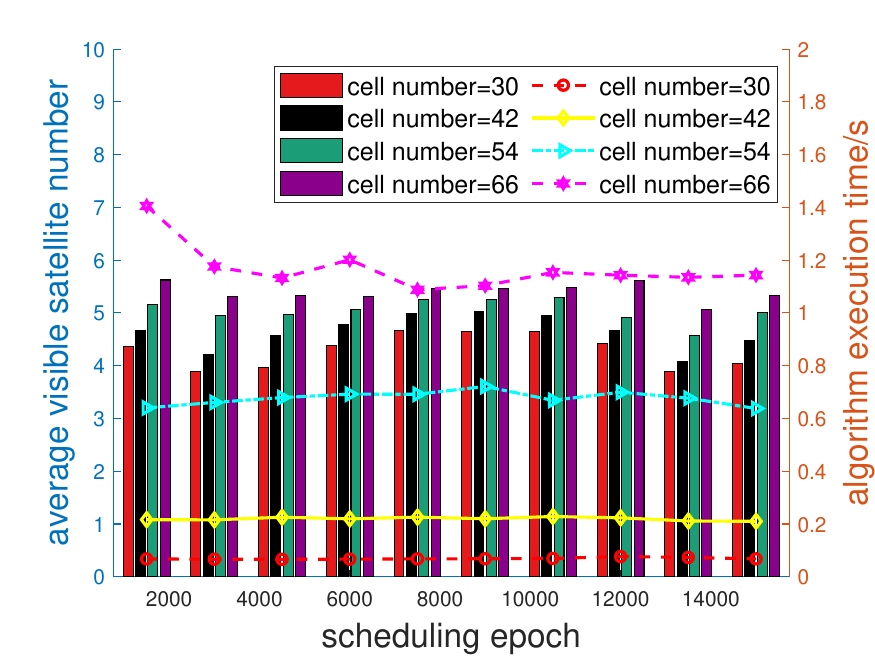}
        }
    \caption{Figure (a) shows the performance of our proposal with different visible satellite numbers under the given cell number, and figure (b) illustrates the performance of our proposal with different cell numbers, where the average visible satellite numbers are represented by bars under different scheduling epochs and solid lines indicate the average execution times.}\label{fig17}
\end{figure}

Finally, to evaluate the scalability of our proposal, we adjust the minimum elevation angle and number of cells, and the simulation results are illustrated in Fig.~\ref{fig17}, where the average visible satellite numbers are represented by bars under different scheduling epochs, and solid lines indicate the average execution time.
According to Fig.~\ref{fig17}(a), it can be observed that the number of visible satellite changes only causes the execution time of our proposal to fluctuate slightly.
Moreover, Fig.~\ref{fig17}(b) indicates that the average execution time rapidly increases with the number of cells.
In conclusion, the simulation results verify our computational complexity analysis in Section~\ref{sec:Complexity Alalysis}: the complexity of our proposal is mainly incurred by cell number, and it has a low correlation with the number of satellites.
Hence, our proposal can be adopted into mega-constellation scenarios.
Moreover, the average execution time of our proposal is larger than the duration of one scheduling epoch.
To control the execution time cost, the cells can be divided into multiple clusters in scenarios with massive cells, and then the beam management plan can be performed separately in each cluster.

\section{Conclusions}\label{sec:con}
In this paper, we have investigated beam management in dynamic LEO satellite networks with moving satellites and random traffic arrival, aiming to achieve a low average beam revisit time and inter-satellite handover frequency.
Facing the challenges incurred by time-averaged terms in the formulated problem and tight coupling among multi-dimensional resource management decisions,
we have divided the primal problem into three subproblems, including serving beam allocation, beam service time allocation and serving satellite allocation.
Under any given serving relationships between satellites and cells, serving beam allocation problem has been converted into a maximum weighted independent set searching problem based on a conflict graph, which has been solved by a low-complexity algorithm.
Subsequently, beam service time allocation algorithm has been designed to reallocate the service time of cells, aiming to balance beam revisit time and network stability.
Finally, we have proposed a serving satellite allocation algorithm to determine the satellite-cell serving relationship.
The superiority of our proposal has been verified through numerical results, which show that our proposal achieves a better average beam revisit time and strong network stability with a low inter-satellite handover frequency compared to benchmark schemes.


\end{document}